\documentclass[12pt]{article}
\pdfoutput=1

\usepackage[utf8]{inputenc}
\usepackage[left=2.55cm, right=2.55cm, top=2.55cm, bottom=2.55cm]{geometry}
\usepackage{amsmath,amssymb,amsbsy}
\usepackage{slashed}
\usepackage{xcolor}
\usepackage{graphicx}
\usepackage{cancel}
\usepackage{url}
\usepackage{cancel}
\usepackage{tcolorbox} 
\usepackage{cite}
\usepackage[colorlinks=true,allcolors=darkpurple,pdfborder={0 0 0},linktocpage=false]{hyperref}
\usepackage{tabularx,booktabs}
\usepackage{multicol}
\usepackage{units}
\usepackage{xspace}
\usepackage[labelfont=bf]{caption}
\usepackage[section]{placeins}
\usepackage{enumitem}
\usepackage{ulem}
\usepackage{subcaption}
\usepackage{multirow}
\usepackage{comment}
\usepackage{verbatim}
\usepackage{ulem}
\useunder{\uline}{\ul}{}
\usepackage{tensor}
\usepackage{array} 
\newcolumntype{C}[1]{>{\centering\arraybackslash}p{#1}}
\renewcommand{\arraystretch}{1.5}
\usepackage[compat=1.1.0]{tikz-feynman}
\tikzfeynmanset{warn luatex=false}

\definecolor{darkred}{rgb}{0.6,0,0}
\definecolor{darkpurple}{rgb}{0.5,0,0.5}
\definecolor{lightblue}{rgb}{0.93, 0.95, 1.0}
\definecolor{mygray}{rgb}{0.9,0.9,0.9}
\definecolor{avblue}{rgb}{0.0, 0.0, 0.8}
\definecolor{javi}{rgb}{0.5,0,1}
\definecolor{asparagus}{rgb}{0.53, 0.66, 0.42}
\definecolor{aqua}{rgb}{0.4, 0.6, 0.7}

\def\abs[#1]{\lvert #1\rvert}
\def\L{\mathcal{L}}
\def\M{\mathcal{M}}

\def\hc{\text{h.c.}}

\def\z2{$\mathbb{Z}_2$}
\def\Tr{\text{Tr}}

\def\id{\mathbb{I}}

\def\U1L{$\mathrm{U(1)}_L$}
\def\3221{$SU(3)_c\times SU(2)_1\times SU(2)_2\times U(1)_Y$}
\def\221{$SU(2)_1\times SU(2)_2\times U(1)_Y$}

\newcommand{\Ss}{{\cal S}}
\newcommand{\Pp}{{\cal P}}
\newcommand{\Hh}{{\cal H}}

\newcommand {\ignore}[1]{}

\newcommand{\AddrIFIC}{%
  Instituto de F\'{i}sica Corpuscular, CSIC-Universitat de Val\`{e}ncia, 46980 Paterna, Spain}

\newcommand{\AddrFISTEO}{%
  Departament de F\'{\i}sica Te\`{o}rica, Universitat de Val\`{e}ncia, 46100 Burjassot, Spain}

\begin{document}


\begin{center}
\vspace*{15mm}

\vspace{1cm}
{\Large \bf 
Scotogenic mechanism from an extended $\boldsymbol{SU(2)_1 \times SU(2)_2 \times U(1)_Y}$ electroweak symmetry
} \\
\vspace{1cm}

{\bf Julio Leite$^{\text{a}}$, Javier Perez-Soler$^{\text{a}}$, Avelino Vicente$^{\text{a,b}}$}

\vspace*{.5cm}
 $^{(\text{a})}$ \AddrIFIC \\\vspace*{.2cm} 
 $^{(\text{c})}$ \AddrFISTEO

\vspace*{.3cm}
\href{mailto:julio.leite@ific.uv.es}{julio.leite@ific.uv.es},
\href{mailto:javier.perez.soler@ific.uv.es}{javier.perez.soler@ific.uv.es},
\href{mailto:avelino.vicente@ific.uv.es}{avelino.vicente@ific.uv.es}
\end{center}

\vspace*{10mm}
\begin{abstract}\noindent\normalsize
We propose an extension of the electroweak sector of the Standard Model in which the gauge group $SU(2)_L$ is promoted to $SU(2)_1 \times SU(2)_2$. This framework naturally includes a viable dark matter candidate and generates neutrino masses radiatively \textit{à la Scotogenic}. Our scenario can be viewed as an ultraviolet extension of the Scotogenic mechanism, addressing some of its shortcomings. The resulting phenomenology may be probed through a range of experimental signatures, from precision electroweak measurements to searches for lepton flavor violation.
\end{abstract}

\section{Introduction}
\label{sec:intro}

There are two central problems in modern physics: the origin and explanation of the tiny neutrino masses required to account for oscillation data, and the nature of the dark matter (DM) that constitutes about 25\% of the Universe's energy content. The existence of neutrino masses has been firmly established by decades of experiments, which have now entered the precision era, while the presence of dark matter is supported by a broad range of astrophysical and cosmological observations, including the cosmic microwave background and galaxy rotation curves, among many others.

These two problems may be unrelated and have entirely independent solutions. In fact, DM could be composed of macroscopic objects, not necessarily within the domain of particle physics. However, it is tempting to consider a common solution to both issues~\cite{Avila:2025qsc}. One of the most popular scenarios that addresses them simultaneously is the Scotogenic model~\cite{Tao:1996vb,Ma:2006km}. In its original formulation, this economical framework extends the Standard Model (SM) particle content by introducing only four additional fields (three fermion singlets and one scalar doublet) and a simple \z2 symmetry. With just these ingredients, it generates radiative Majorana neutrino masses and provides a viable DM candidate. Although many variants of the Scotogenic model exist, they all share a common feature: simplicity. For a review of radiative neutrino mass models, including the Scotogenic model and its variants, see~\cite{Cai:2017jrq}.

There are, however, some elements of the Scotogenic model that are not entirely satisfactory:
\begin{itemize}
\item The \z2 parity in the Scotogenic model is an ad hoc symmetry. While this is a legitimate model-building approach, it is compelling to explore extensions of the Scotogenic model that either provide a dynamical origin for the \z2 symmetry or allow it to emerge automatically from an underlying gauge symmetry.
\item In Scotogenic scenarios, it is common to assume that lepton number is only slightly broken. This is typically implemented by introducing a small coupling in the scalar potential, usually denoted by $\lambda_5$. In the limit $\lambda_5 \to 0$, lepton number is recovered. While the smallness of this parameter is technically natural~\cite{tHooft:1979rat}, its origin remains unexplained within the Scotogenic model itself.
\end{itemize}

These shortcomings of the Scotogenic model can be addressed in extended scenarios featuring larger symmetries and additional energy scales. For example, a natural origin for the \z2 symmetry is a global lepton number symmetry, $U(1)_L$, which is properly broken to leave a remnant \z2. Moreover, the smallness of the lepton number–violating parameter $\lambda_5$ can be attributed to its effective origin, arising after integrating out a heavy degree of freedom. These ideas have been explored in detail in~\cite{Escribano:2021ymx,Portillo-Sanchez:2023kbz}.

In this work, we consider the extended electroweak symmetry $SU(2)_1 \times SU(2)_2 \times U(1)_Y$. We show that a suitable choice of representations under both $SU(2)$ groups leads to a Scotogenic scenario once the symmetry is spontaneously broken to $SU(2)_L \times U(1)_Y$. All the attractive features of the original Scotogenic model are preserved: our setup generates radiative neutrino masses and includes a viable DM candidate. In addition, the extended electroweak symmetry addresses the aforementioned shortcomings of the Scotogenic model. In our framework, the \z2 parity is neither imposed nor tied directly to lepton number. Instead, it emerges as an accidental symmetry. We also introduce a scalar transforming as a triplet under $SU(2)_2$. Assuming its mass lies well above any other scale in the model, one naturally obtains a small explicit violation of lepton number and light Majorana neutrinos.

Two recent works that share a similar spirit to ours are \cite{VanDong:2023xmf} and \cite{CarcamoHernandez:2025eyt}. However, their realizations are completely different. In~\cite{VanDong:2023xmf}, the Scotogenic model is also recovered after extending the electroweak group, specifically to $SU(3)_L \times U(1)_X \times U(1)_G$. The authors of~\cite{CarcamoHernandez:2025eyt} also promote $SU(2)_L$ to $SU(2)_1 \times SU(2)_2$, but with a strongly coupled $SU(2)_2$. Moreover, the origin of the small violation of lepton number is not addressed in either of these works.

The rest of the paper is organized as follows: in Sec.~\ref{sec:model} we introduce our model, describe its basic features and derive the effective theory obtained at low energies after integrating out the heavy scalar triplet. The resulting particle spectrum is studied in detail in Sec.~\ref{sec:spectrum}. In particular, and among other analytical results, in this Section we compute the 1-loop neutrino mass matrix in our model. Section~\ref{sec:pheno} is devoted to the phenomenology of our model. We obtain limits to the relevant energy scales and discuss lepton flavor violation and dark matter. Finally, we summarize and conclude in Sec.~\ref{sec:conclu}. Some details about $SU(2)$ are left for Appendix~\ref{app:SU2stuff}.
\section{The model}
\label{sec:model}

Our model promotes the electroweak gauge group to \221, with the SM $SU(2)_L$ contained in the double $SU(2)$ product. The gauge bosons and couplings of the extended electroweak group are denoted as
\begin{align}
\begin{aligned}
SU(2)_1 &: \quad g_1,\quad W_1^a\, , \\
SU(2)_2 &: \quad g_2,\quad W_2^a \, , \\
U(1)_Y &: \quad g^\prime,\quad B\,, \\
\end{aligned}
\end{align}
where $a=1,2,3$ is an $SU(2)$ triplet index, and Lorentz indices were omitted. The SM left-handed fermions are assumed to transform as doublets under the first $SU(2)$ factor, i.e., 
\begin{align}
\begin{aligned}
Q &= \left( {\bf 3} , {\bf 2} , {\bf 1} \right)_{\frac{1}{6}} \, ,  &   L &= \left( {\bf 1} , {\bf 2} , {\bf 1} \right)_{-\frac{1}{2}} \, , \\
u_R &= \left( {\bf 3} , {\bf 1} , {\bf 1} \right)_{\frac{2}{3}} \, ,  &   e_R &= \left( {\bf 1} , {\bf 1} , {\bf 1} \right)_{-1} \, , \\
d_R &= \left( {\bf 3} , {\bf 1} , {\bf 1} \right)_{-\frac{1}{3}} \, , &
\end{aligned}
\end{align}
where we denote the representations under ($SU(3)_c$, $SU(2)_1$, $SU(2)_2$)$_{U(1)_Y}$. The SM doublets can be
decomposed in $SU(2)_1$ components in the usual way,
\begin{align}
\begin{aligned}
Q & =
\begin{pmatrix}
u \\
d 
\end{pmatrix}_L
\, , \qquad
L =
\begin{pmatrix}
\nu \\
e 
\end{pmatrix}_L \, .
\end{aligned}
\end{align}
In addition, we introduce 2 generations of the bidoublet of right-handed fermions $\chi$ transforming as $\left( {\bf 1} , {\bf 2} , {\bf \bar 2} \right)_0$. We choose it to be a real bidoublet of $SU(2)_1 \times SU(2)_2$,~\footnote{In our treatment of the fermion bidoublet $\chi$ we follow the conventions of~\cite{Garcia-Cely:2015quu}.}
\begin{equation}
\chi \equiv \chi_R= \begin{pmatrix}
\chi_0 & \chi^+\\
\chi^- &  -\chi_0^c
\end{pmatrix}_R \,
\end{equation}
with $\chi^- = \left(\chi^+\right)^*$ and where row and column indices are $SU(2)_1$ and $SU(2)_2$ indices, respectively. The fact that $\chi_R$ is real implies that $\chi \equiv \chi_R$ and its left-handed equivalent\footnote{Real representations are usually discussed when dealing with scalar multiplets, which are Lorentz singlets and thus we get $\phi=\Tilde \phi$ without a chirality flip. In the case of a fermion, such as $\chi$, one can get a real multiplet in the sense of $\Psi=\Tilde \Psi$ with $\Psi = \Psi_L\oplus \Psi_R$, that is, the bispinor.} are related via the dual $\chi_L=\Tilde{\chi_R} = C \, (\chi_R)^c \, C$, with $C$ being the $SU(2)$ conjugation matrix in the doublet representation (see Appendix~\ref{app:SU2stuff} for details on $SU(2)$). This duality relation implies that $\chi_R$ and $\chi_L$ transform in the same way under the gauge group, differing only in their Lorentz group representation. In particular, it enforces their lepton numbers (and any other $U(1)$ charges) to be zero, since a non-zero lepton number would make that of the bispinor $\Psi=\chi_L+\chi_R$ undefined. It also implies that the $\chi$ bidoublets do not contribute to gauge anomalies. We will from now on write $\chi_R\equiv\chi$ and $\chi_L\equiv\Tilde\chi$.

The scalar sector of the model is composed by four multiplets. In addition to the pair of doublets $H$ and $\eta$, each of them charged under one of the $SU(2)$ factors, it includes the triplet $\Delta$ and the bitriplet $\Omega$, i.e.,
\begin{align}
\begin{aligned}
H &= \left( {\bf 1} , {\bf 2} , {\bf 1} \right)_{\frac{1}{2}} \,, \qquad 
\eta &= \left( {\bf 1} , {\bf 1} , {\bf 2} \right)_{\frac{1}{2}} \,, \\
\Delta &= \left( {\bf 1} , {\bf 1} , {\bf 3} \right)_{1} \,, \qquad
\Omega &= \left( {\bf 1} , {\bf 3} , {\bf \bar 3} \right)_{0} \, .
\end{aligned}
\end{align}
The $\Omega$ bitriplet is assumed to be real. The scalar multiplets can be decomposed in terms of their $SU(2)_1 \times SU(2)_2$ components as
\begin{align}\label{eq:scalarMultiplets1}
    H = \left(\begin{array}{c}
         H^+  \\
         H_0 
    \end{array}\right)\,,\quad
    \eta = \left(\begin{array}{c}
         \eta^+  \\
         \eta_0 
    \end{array}\right)
\end{align}
and
\begin{align}\label{eq:scalarMultiplets2}
\Delta = \left(\begin{array}{c}
         \Delta^{++}  \\
         \Delta^+ \\
         \Delta_0
    \end{array}\right)\,,\quad
    \Omega = \left(\begin{array}{ccc}
        \Omega_0 & \Omega_1^+ & \Omega^{++} \\
        -\Omega_2^- & \xi_0 & \Omega_2^+ \\
        \Omega^{--} & -\Omega_1^- & \Omega_0^*
    \end{array}\right) \, .
\end{align}
Here $\Omega_{1,2}^- = \left( \Omega_{1,2}^+ \right)^*$ and $\Omega^{--} = \left( \Omega^{++} \right)^*$. Similarly, in the following we will denote $H^- = \left(H^+\right)^*$, $\eta^- = \left(\eta^+\right)^*$, $\Delta^- = \left(\Delta^+\right)^*$ and $\Delta^{--} = \left(\Delta^{++}\right)^*$. We note that $\Omega$ being self-dual enforces its $(2,2)$ component to be a real field, $\xi_0^* = \xi_0$. Again, row and column indices in $\Omega$ are $SU(2)_1$ and $SU(2)_2$ indices, respectively. The leptonic and scalar particle content of the model, as well as their representations under the extended electroweak group, is shown in Tab.~\ref{table:particleContent}. 

\setlength{\tabcolsep}{1em} 
\renewcommand{\arraystretch}{1.5}
\begin{table}[t!]
\centering
\begin{tabular}[t!]{||c c c c c||} 
 \hline
 Field & Generations & $SU(2)_1$ & $SU(2)_2$ & $U(1)_Y$ \\ [0.5ex] 
 \hline\hline
 $L$ & 3 & $\mathbf{2}$ & $\mathbf{1}$ & $-1/2$ \\ 
 $e_R$ & 3 & $\mathbf{1}$ & $\mathbf{1}$ & $-1$ \\
 $\chi$ & 2 & $\mathbf{2}$ & $\overline{\mathbf{2}}$ & $0$ \\ [1ex] 
 \hline\hline
 $H$ & 1 & $\mathbf{2}$ & $\mathbf{1}$ & $1/2$ \\
 $\eta$ & 1 & $\mathbf{1}$ & $\mathbf{2}$ & $1/2$ \\
 $\Delta$ & 1 & $\mathbf{1}$ & $\mathbf{3}$ & $1$ \\ 
 $\Omega$ & 1 & $\mathbf{3}$ & $\overline{\mathbf{3}}$ & $0$ \\[1ex] 
 \hline
\end{tabular}
\vspace{3mm}
\caption{Leptonic and scalar particle content of the model.
\label{table:particleContent}}
\end{table}

\subsubsection*{Yukawa interactions and scalar potential}

The SM quarks have Yukawa couplings with the $H$ doublet in the usual way. Regarding the SM leptons, in addition to their usual Yukawa coupling, new terms are allowed by the gauge symmetry of the model,
\begin{align}  \label{eq:LY}
-\mathcal{L}_{\ell} =  y \, \overline{L} \, H \, e_R + Y \, \tilde{\eta}^\dagger \, \overline{\chi} \, L+ \frac{M}{2} \, \Tr{\left[\,\overline{\tilde{\chi}}\chi\,\right]} + y_\chi\Tr\left[{\overline{\tilde{\chi}}\; \tau_a\; \chi\;\tau^b}\right](\Omega^\dagger)\indices{^b_a}+ \hc \; ,
\end{align}
where  $\tau^a$ are the $SU(2)$ generators (we use lower index for their Hermitian conjugates) for the doublet representation in the spherical basis. We refer to Appendix~\ref{app:SU2stuff} for details. We also note that the transpose in the Dirac conjugate also acts on the $SU(2)_1\times SU(2)_2$ indices. The couplings $y$ and $Y$ are two general Yukawa matrices, $3 \times 3$ and $3 \times 2$, respectively, while $y_\chi$ and $M$ are two $2\times 2$ symmetric matrices. Without loss of generality, we will assume $y$ and $M$ to be diagonal.

The scalar potential of the model can be written as
\begin{align}
\mathcal V &= m_H^2 \, H^\dagger H + m_\eta^2 \, \eta^\dagger \eta + m_\Delta^2 \, \Delta^\dagger \Delta + m_\Omega^2 \, \Tr \left( \Omega^\dagger\Omega \right) \nonumber \\
&+ \frac{\lambda_H}{2} \, (H^\dagger H)^2 + \frac{\lambda_\eta}{2} \, (\eta^\dagger \eta)^2 + \frac{\lambda_{\Delta 1}}{2} \, (\Delta^\dagger \Delta)^2 + \frac{\lambda_{\Delta 2}}{2} \, (\Delta^\dagger t^a \Delta)(\Delta^\dagger t_a \Delta) + \frac{\lambda_{\Omega 1}}{2} \, \left[ \Tr \left( \Omega^\dagger\Omega \right)\right]^2 \, \nonumber\\
&+ \frac{\lambda_{\Omega 2}}{2} \, \Tr \left(\Omega^\dagger\Omega\,\Omega^\dagger\Omega \right)+ \lambda_{H\eta} \, (H^\dagger H)(\eta^\dagger \eta) + \lambda_{H\Delta} \, (H^\dagger H) (\Delta^\dagger \Delta) + \lambda_{H\Omega} \, (H^\dagger H) \, \Tr \left( \Omega^\dagger \Omega \right) \nonumber \\
&+ \lambda_{\eta\Delta 1} \, (\eta^\dagger \eta) (\Delta^\dagger \Delta) + \lambda_{\eta\Delta 2} \, (\eta^\dagger\tau^a\eta)(\Delta^\dagger t_a\Delta) + \lambda_{\eta\Omega} \, (\eta^\dagger \eta) \, \Tr \left( \Omega^\dagger \Omega \right) + \lambda_{\Delta\Omega} \, (\Delta^\dagger \Delta) \, \Tr \left( \Omega^\dagger \Omega \right) \nonumber \\
&+ \mu_1 \, \Tr \left[ \Omega^\dagger t_a \; \Omega \; t^b \right](\Omega^\dagger)\indices{^b_a} + \left[ \mu_2 \, \Delta^a(\eta^\dagger \tau^a \Tilde{\eta}) +  \lambda_{H\Delta\Omega} \,  (H^\dagger \tau^a \Tilde{H})\Omega^{ab}\Delta^b+\hc \right] \, . \label{eq:potential}
\end{align}
Here $\tau^a$ and $t^a$ are the $SU(2)$ generators for the doublet and triplet representations, respectively, in the spherical basis. Again, we use lower index to denote their Hermitian conjugates and refer to Appendix~\ref{app:SU2stuff} for further details. Moreover, the parameters $m_i^2$, with $i=H,\eta,\Delta,\Omega$ have dimensions of mass$^2$, the parameters $\mu_i$, with $i=1,2$, have dimensions of mass and the rest are dimensionless.

By inspecting the Yukawa Lagrangian in Eq.~\eqref{eq:LY} and the scalar potential in Eq.~\eqref{eq:potential}, it is easy to realize that lepton number is explicitly broken by the simultaneous presence of the last two terms of the scalar potential. In fact, if either $\mu_2$ or $\lambda_{H\Delta\Omega}$ vanish, a definition of a conserved $U(1)_\ell$ lepton number would be possible, with $\ell(L) = \ell(e_R) = \ell(\chi) = 1$ and $\ell(H) = \ell(\eta) = \ell(\Omega) = 0$. This is the motivation behind the introduction of the $SU(2)_2$ triplet $\Delta$ in the model.

\subsubsection*{Symmetry breaking}

The study of the detailed structure of the complicated scalar potential in Eq.~\eqref{eq:potential} is beyond the scope of our work. Instead, we will simply assume that its minimum is characterized by the vacuum expectation values (VEVs)
\begin{align}\label{eq:VEVs}
\langle H \rangle = \frac{1}{\sqrt{2}}
\begin{pmatrix} 
0 \\
v_H \end{pmatrix},
\quad \quad 
\langle \eta \rangle = 0 \, ,
\quad \quad
\langle \Delta \rangle = \frac{1}{\sqrt{2}}
\begin{pmatrix} 
0 \\
0 \\
v_\Delta \end{pmatrix},
\quad \quad 
\langle \Omega \rangle = \frac{1}{8}
\begin{pmatrix}
v_\Omega & 0 & 0 \\
0 & v_\xi & 0 \\
0 & 0 & v_\Omega
\end{pmatrix} \, .
\end{align}
Assuming the VEV hierarchy
\begin{equation}\label{eq:VEVhier}
v_\Delta \ll v_H \ll v_\Omega, v_\xi \, ,
\end{equation}
the spontaneous breaking of the electroweak symmetry proceeds as
\begin{align} \label{eq:SSB}
SU(2)_1 \times SU(2)_2 \times U(1)_Y \stackrel{v_\Omega, v_\xi}{\longrightarrow} SU(2)_L \times U(1)_Y \stackrel{v_H, v_\Delta}{\longrightarrow} U(1)_{\rm em}\,,
\end{align}
with $v = \sqrt{v_H^2 + 2 v_\Delta^2} \approx v_H \approx 246$ GeV the usual electroweak VEV. In the first step, the $SU(2)_1 \times SU(2)_2$ group gets broken by the non-zero VEV of $\Omega$, which preserves a diagonal subgroup that we identify with the SM $SU(2)_L$. Under this group, $H$ and $\Delta$ transform as a doublet and as a triplet, respectively. Their VEVs break $SU(2)_L \times U(1)_Y$ to a remnant $U(1)_{\rm em}$. The electric charge operator, which is the generator of this unbroken symmetry, is then given by
\begin{equation}
Q = \left( T_3^1+T_3^2 \right) + Y = T_3^L + Y \, ,
\end{equation}
with $T_3^a$ the diagonal generator of $SU(2)_a$. The VEV configuration in Eq.~\eqref{eq:VEVs} enforces four tadpole equations,
\begin{equation}
    \frac{\partial \mathcal{V}}{\partial \phi_i} = 0 \, , \quad \text{with} \,\, \phi_i = \left\{ H_0, \Delta_0, \Omega_0, \xi_0 \right\} \, .
\end{equation}
In the minimum of the scalar potential they read
\begin{align}
64 \, m_H^2 \, v_H &+ 32 \, \lambda_H \, v_H^3+32 \, \lambda_{H\Delta} \, v_H \, v_\Delta^2+ \lambda_{H\Omega} \, v_H \, (v_\xi^2+2v_\Omega^2) +8 \, \lambda_{H\Delta\Omega}^* \, v_H \, v_\Delta \, v_\Omega=0\, , \label{eq:tadH} \\
64 \, m_\Delta^2 \, v_\Delta &+ 32 \, (\lambda_{\Delta1}+\lambda_{\Delta2}) \, v_\Delta^3 + 32 \, \lambda_{H\Delta} \, v_\Delta \, v_H^2 + \lambda_{\Delta\Omega} \, v_\Delta \left( 2 \, v_\Omega^2 +  v_\xi^2 \right) \nonumber \\
&+ 4 \, \lambda_{H\Delta\Omega} \, v_H^2 \, v_\Omega = 0\, , \label{eq:tadDelta} \\
64 \, m_\Omega^2 \, v_\Omega &+ (2\lambda_{\Omega1}+\lambda_{\Omega2}) \, v_\Omega^3 + 32 \, \lambda_{H\Omega} \, v_\Omega \, v_H^2 + 32 \, \lambda_{\Delta\Omega} \, v_\Omega \, v_\Delta^2 + \lambda_{\Omega1} \, v_\Omega \, v_\xi^2 + 24 \, \mu_1 \, v_\Omega \, v_\xi \nonumber \\
&+ 64 \, \lambda_{H\Delta\Omega}^* \, v_H^2 v_\Delta =0\, , \label{eq:tadOmega} \\
64 \, m_\Omega^2 \, v_\xi &+ (\lambda_{\Omega1}+\lambda_{\Omega2}) \, v_\xi^3 + 32 \, \lambda_{H\Omega} \, v_\xi \, v_H^2 + 32 \, \lambda_{\Delta\Omega} \, v_\xi \, v_\Delta^2 + 2 \, \lambda_{\Omega1} \, v_\xi \, v_\Omega^2 + 24 \, \mu_1 \, v_\xi^2 =0 \, . \label{eq:tadXi}
\end{align}
Our assumption of a small triplet VEV $v_\Delta$, see Eq.~\eqref{eq:VEVhier}, motivates two observations:
\begin{itemize}
\item Eqs.~\eqref{eq:tadOmega} and \eqref{eq:tadXi} correspond to the tadpoles of $\Omega_0$ and $\xi_0$, respectively. It is easy to check that the last term in Eq.~\eqref{eq:tadOmega} precludes $v_\Omega = v_\xi$ from being a valid vacuum of our model. We notice that this condition would ensure that a remnant $SU(2)_L$ diagonal subgroup is conserved after $SU(2)_1 \times SU(2)_2$ breaking. Therefore, this is completely consistent with our assumptions. In the presence of $v_\Delta \neq 0$, the vacuum that preserves $SU(2)_L$ is no longer possible. This is more easily seen by solving Eqs.~\eqref{eq:tadH}-\eqref{eq:tadXi} for the squared mass parameters $m_H^2$, $m_\Delta^2$, $m_\Omega^2$ and $v_\Delta$. One finds the condition
\begin{equation} \label{eq:vDelta}
v_\Delta = \frac{v_\Omega}{64 \, \lambda_{H\Delta\Omega}^* \, v_H^2 \, v_\xi} \, (v_\Omega^2-v_\xi^2)(24 \, \mu_1 - \lambda_{\Omega2} \, v_\xi) \, .
\end{equation}
We conclude that $v_\Omega \neq v_\xi$ due to $v_\Delta \neq 0$. However, the VEV hierarchy in Eq.~\eqref{eq:VEVhier} motivates the choice $v_\Omega \approx v_\xi$. More precisely, the difference $v_\Omega^2-v_\xi^2$ will be proportional to $v_\Delta$ and then small, thus implying an approximate $SU(2)_L$ symmetry after the bitriplet $\Omega$ acquires a non-zero VEV.
\end{itemize}
\begin{itemize}
\item The hierarchy in Eq.~\eqref{eq:VEVhier} can be used to obtain an approximate solution to Eq.~\eqref{eq:tadDelta}. At leading order in $v_\Delta$ one finds
\begin{equation} \label{eq:oomDelta}
\frac{1}{m_\Delta^2} \approx - \frac{16 \, v_\Delta}{\lambda_{H\Delta\Omega} \, v_H^2 \, v_\Omega} \, .
\end{equation}
Therefore, $v_\Delta \to 0$ is naturally obtained when $m_\Delta \to \infty$. When the triplet mass is well above all the other mass scales in the model, its VEV becomes small, as assumed in Eq.~\eqref{eq:VEVhier}. This justifies the construction of a low-energy theory in which $\Delta$ is integrated out.
\end{itemize}

\subsubsection*{Low-energy theory}

In the following, we consider $m_\Delta$ to be much larger than any other energy scale in the model. Under this assumption, one can integrate out $\Delta$ and determine a low-energy scalar potential valid at energy scales below $m_\Delta$, where $\Delta$ is absent but new operators with dimensions higher than 4 appear multiplied by the corresponding powers of $1/m_\Delta$. The result, obtained at tree-level by following the method of Ref.~\cite{de_Blas_2015}, is
\begin{align}
\mathcal{V}_\text{eff} &= m_H^2 \, H^\dagger H + m_\eta^2 \, \eta^\dagger \eta + m_\Omega^2 \, \Tr \left( \Omega^\dagger\Omega \right) \, \nonumber\\ & + \frac{\lambda_H}{2} \, (H^\dagger H)^2 +\left(\frac{\lambda_\eta}{2} + \frac{|\mu_2|^2}{m_\Delta^2}\right) \, (\eta^\dagger \eta)^2 + \frac{\lambda_{\Omega 1}}{2} \, \left[ \Tr \left( \Omega^\dagger\Omega \right)\right]^2 \, \nonumber\\ & + \frac{\lambda_{\Omega 2}}{2}\Tr \left(\Omega^\dagger\Omega\,\Omega^\dagger\Omega \right)+ \lambda_{H\eta} \, (H^\dagger H)(\eta^\dagger \eta) + \lambda_{H\Omega} \, (H^\dagger H) \, \Tr \left( \Omega^\dagger \Omega \right) \nonumber + \lambda_{\eta\Omega} \, (\eta^\dagger \eta) \, \Tr \left( \Omega^\dagger \Omega \right) \nonumber \\
&+ \mu_1 \, \Tr \left[ \Omega^\dagger t_a \; \Omega \; t^b \right](\Omega^\dagger)\indices{_b_a} +  \frac{2|\lambda_{H\Delta\Omega}|^2}{m_\Delta^2} \,  (H^\dagger \tau^b \Tilde{H}) \, \Omega_{ba}\Omega^\dagger_{ac} \,(\Tilde{H}^\dagger \tau_c H)\nonumber\\
&+\left[\frac{2\mu_2^*\lambda_{H\Delta\Omega}}{m_\Delta^2}(H^\dagger \tau^a \Tilde{H}) \, \Omega_{ab} \, (\Tilde{\eta}^\dagger \tau_b \eta)  + \hc \right]
+ \mathcal{O}(1/m_\Delta^4)\;. \label{eq:Veff}
\end{align}

Again, we will assume the minimum
\begin{align}\label{eq:VEVs}
\langle H \rangle = \frac{1}{\sqrt{2}}
\begin{pmatrix} 
0 \\
v_H \end{pmatrix},
\quad \quad 
\langle \eta \rangle = 0 \, ,
\quad \quad 
\langle \Omega \rangle = \frac{1}{8}
\begin{pmatrix}
v_\Omega & 0 & 0 \\
0 & v_\xi & 0 \\
0 & 0 & v_\Omega
\end{pmatrix} \, .
\end{align}
This VEV configuration leads to three non-trivial tadpole equations. When evaluated at the minimum of the potential, they read
\begin{align}
64 \, m_H^2 \, v_H &+ 32 \, \lambda_H \, v_H^3 + \lambda_{H\Omega} \, v_H \, (v_\xi^2+2v_\Omega^2) + \frac{|\lambda_{H\Delta\Omega}|^2 \, v_H^3 \, v_\Omega^2}{m_\Delta^2} = 0 \, , \label{eq:tadHEFT} \\
64 \, m_\Omega^2 \, v_\Omega &+ (2\lambda_{\Omega1}+\lambda_{\Omega2}) \, v_\Omega^3 + 32 \, \lambda_{H\Omega} \, v_\Omega \, v_H^2  + \lambda_{\Omega1} \, v_\Omega \, v_\xi^2 + 24 \, \mu_1 \, v_\Omega \, v_\xi + \frac{8 \, |\lambda_{H\Delta\Omega}|^2 \, v_H^4}{m_\Delta^2} = 0\, , \label{eq:tadOmegaEFT} \\
64 \, m_\Omega^2 \, v_\xi &+ (\lambda_{\Omega1}+\lambda_{\Omega2}) \, v_\xi^3 + 32 \, \lambda_{H\Omega} \, v_\xi \, v_H^2 + 2 \, \lambda_{\Omega1} \, v_\xi \, v_\Omega^2 + 24 \, \mu_1 \, v_\xi^2 = 0 \, . \label{eq:tadXiEFT}
\end{align}

Again, we emphasize that the configuration $v_\Omega = v_\xi$ is not compatible with $v_\Delta \neq 0$. In the case of the effective theory, this translates into Eqs.~\eqref{eq:tadOmegaEFT} and \eqref{eq:tadXiEFT} not being compatible with $v_\Omega = v_\xi$ unless the limit $m_\Delta \to \infty$ is taken.

We note that the Lagrangian of our model contains an accidental \z2 symmetry, under which $\chi$ and $\eta$ are odd, while the rest of the fields are even. Indeed, one can easily check in Eqs.~\eqref{eq:LY} and \eqref{eq:potential} that $SU(2)_1 \times SU(2)_2$ enforces these two fields to always appear in pairs (individually or combined).~\footnote{One may naively think that the \z2 symmetry is a subgroup of $SU(2)_1 \times SU(2)_2$, but this is not the case. Although the \z2 symmetry is enforced by this product of $SU(2)$'s, it also relies on our choice of scalar representations. For instance, it may have seemed preferable to choose a bidoublet representation for $\Omega$, instead of the larger bitriplet representation introduced in our model. However, in such case, a trilinear $H^\dagger \Omega \, \eta$ term would be allowed in the scalar potential, hence breaking the accidental \z2 of the model.} This symmetry is not broken by the scalar VEVs in Eq.~\eqref{eq:VEVs} and remains exact at all scales. This is crucial for the particle spectrum and phenomenology of the model, as explained below.

Finally, once $\Omega$ acquires a VEV an effective $\displaystyle \frac{\lambda_5}{2} \left(H^\dagger \eta \right)^2$ quartic term is generated, with
\begin{equation} \label{eq:lam5}
   \lambda_5 = \frac{v_\Omega \, \lambda_{H\Delta\Omega} \, \mu_2^*}{4 \,m_\Delta^2}\;.
\end{equation}
The presence of this coupling in the low-energy theory leads to lepton number violation. Since $v_\Omega \, \mu_2 \ll m_\Delta^2$, we find that all lepton number breaking effects, including Majorana neutrino masses, are naturally suppressed by $\lambda_5 \ll 1$~\cite{Escribano:2021ymx,Portillo-Sanchez:2023kbz}.
\section{Particle spectrum}
\label{sec:spectrum}

In this Section we provide detailed analytical results for the masses of the gauge bosons, scalars and fermions in the model. We will work in the effective theory obtained after the triplet $\Delta$ is integrated out.

\subsection{Gauge boson masses}
\label{subsec:gbmasses}

The masses of the gauge bosons originate from the scalar kinetic terms,
\begin{equation}\label{eq:scalarKineticLagrangian}
    \mathcal{L}_{sK} = (D_\mu H)^\dagger (D^\mu H) + (D_\mu \eta)^\dagger (D^\mu \eta) + \Tr\Big\{(D_\mu \Omega)^\dagger (D^\mu \Omega)\Big\}\;, 
\end{equation}
which after SSB lead to terms quadratic in the gauge bosons. The covariant derivatives act on the scalars according to their representations,
\begin{align}
    & H = (\mathbf{1},\mathbf{2},\mathbf{1})_\frac{1}{2}&& \xrightarrow{\quad} && D H = \partial H - ig_1 (\tau_1^a W_1^a) H - ig_Y\frac{1}{2}B H\;,\\[3mm]
    & \eta = (\mathbf{1},\mathbf{1},\mathbf{2})_\frac{1}{2}&& \xrightarrow{\quad} &&D \eta = \partial \eta - ig_2 (\tau_2^a W_2^a) \eta - ig_Y\frac{1}{2}B \eta\;,\nonumber\\[4mm]
    & \Omega = (\mathbf{1},\mathbf{3},\overline{\mathbf{3}})_0&& \xrightarrow{\quad} &&D \Omega = \partial \Omega - ig_1 (t_1^a W_1^a) \Omega + ig_2 \Omega \left[(t_2^*)^a W_2^a\right]\;,\nonumber
\end{align}
where we skip writing the Lorentz index. It is now common to write the $SU(2)$ generators in the cartesian basis, with the gauge bosons given by the triplet of real fields $(W^1,W^2,W^3)$. If we instead choose the spherical basis, the gauge bosons triplet becomes the usual real triplet with its elements given by electric charge eigenstates. As shown in Appendix~\ref{app:SU2stuff}, the spherical basis is related to the cartesian basis via
\begin{align} 
    \left(\begin{array}{c}
         W_r^+  \\
         W_r^0  \\
         -W_r^-
    \end{array}\right) = \frac{1}{\sqrt{2}}\left(\begin{array}{ccc}
        -1 & i & 0 \\
        0 & 0 & \sqrt{2} \\
        1 & i & 0 
    \end{array}\right)\left(\begin{array}{c}
         W_r^1  \\
         W_r^2  \\
         W_r^3
    \end{array}\right)\;,\qquad r=1,2\;.
\end{align}
Note that $W^0 = W^3$ remains a real field. The spherical basis proves to be more useful and intuitive when analyzing terms in the expanded Lagrangian, so we choose it for the rest of the paper (contractions are, of course, basis independent). After SSB, the scalar kinetic Lagrangian in Eq.~\eqref{eq:scalarKineticLagrangian} can be written in terms of the mass matrices for the neutral and charged bosons,
\begin{align}
    \mathcal{L}_0 \equiv \mathcal{L}_{sK}\Big|_\text{Vacuum} = V_n^\dagger \mathcal{M}_n V_n + V_c^\dagger \mathcal{M}_c V_c + \text{additional terms,}
\end{align}
where $V_n$ and $V_c$ contain the neutral and charged bosons, respectively. The mass matrices are easily computed by taking second derivatives of $\mathcal{L}_0$ with respect to the gauge bosons,
\begin{align}
    \mathcal{M}_{ij} = \frac{\partial^2 \mathcal{L}_0}{\partial V_i^* \partial V_j} \;.
\end{align}

\subsubsection*{Neutral gauge bosons}


Our SSB scheme breaks $SU(2)_1\times SU(2)_2\times U(1)_Y$ to $U(1)_\text{em}$ via the bitriplet VEV $\langle \Omega \rangle$ and the usual Higgs VEV $\langle H \rangle$. Therefore, we expect to get two massive bosons $Z_{12}$ and $Z_L$, coming from the breaking of $SU(2)_1\times SU(2)_2$ and $SU(2)_L$, and the massless photon due to the remaining conserved $U(1)_\text{em}$ symmetry. Choosing $V_n =(W^3_1,W^3_2,B)$ as basis (note that these are real fields, so $V_n^* = V_n$), the neutral gauge bosons' mass matrix becomes~\footnote{The $v_\xi$ VEV does not participate in the neutral gauge boson sector. This is due to the fact that the covariant derivative acting on the bitriplet $\Omega$ does not generate any term proportional to $\xi_0$ that involves any of the neutral gauge bosons, only the charged ones. This is analogous to the role played by the $({\bf 1},{\bf 3})_0$ scalar triplet in $SU(2)_L \times U(1)_Y$. Its VEV is known to contribute to the masses of the charged gauge bosons, but not to those of the neutral ones.}
\begin{align}\label{eq:massMatrixNeutralsFirst}
    \mathcal{M}_n = \frac{1}{16}\left(\begin{array}{ccc}
         g_1^2(4v_H^2+v_\Omega^2) & -g_1 g_2 v_\Omega^2 & -4g_1g_Y v_H^2  \\
          -g_1 g_2 v_\Omega^2 & g_2^2 v_\Omega^2 & 0  \\
         -4g_1g_Y v_H^2 & 0 & 4g_Y^2 v_H^2
    \end{array}\right)\, .
\end{align}
It is easy to check that this matrix has one vanishing eigenvalue, as expected. As in the SM, we want to go to a basis where the photon appears explicitly. This can be achieved by doing a couple of rotations; the first one turns $(W_1^3, W_2^3)$ into $(Z_h, W_L^3)$, with $W_L^3$ being the usual $SU(2)_L$ neutral boson. To find such a rotation, we evaluate the scalar kinetic Lagrangian from Eq.~\eqref{eq:scalarKineticLagrangian} in the $SU(2)_L$-conserving limit, defined by $v_\Omega = v_\xi$ and $v_H = 0$. We remind the reader that $v_\Omega = v_\xi$ requires the limit $m_\Delta \to \infty$. The resulting mass matrix is just Eq.~\eqref{eq:massMatrixNeutralsFirst} with $v_H \xrightarrow{} 0$, which has two zero eigenvalues. The non-trivial zero eigenvalue corresponds to $W_L^3$, since $SU(2)_L$ remains unbroken in the first SSB step. The normalized eigenvectors are
\begin{align}
    &Z_h = \frac{1}{\sqrt{g_1^2+g_2^2}}\left(g_1 W_1^3-g_2W_2^3\right)\;,&&
    W_L^3 = \frac{1}{\sqrt{g_1^2+g_2^2}}\left(g_2 W_1^3+g_1W_2^3\right)\;.
\end{align}
The value of the $SU(2)_L$ coupling, $g$, can be determined by inspecting Eq.~\eqref{eq:scalarKineticLagrangian} after replacing the original gauge bosons $(W_1^3, W_2^3)$ by $(Z_h, W_L^3)$.~\footnote{For instance, the coupling of two Higgs doublets with two $W_L^3$ gauge bosons must be $\frac{1}{4} \, g^2$.} It is found to be
\begin{equation} \label{eq:g}
g = \frac{g_1 \, g_2}{\sqrt{g_1^2+g_2^2}} \, .
\end{equation}
This equation introduces a constraint between $g_1$ and $g_2$, leaving one free parameter. Below we will consider the particular case of equal $SU(2)$ gauge couplings, $g_1 = g_2 = \sqrt{2} \, g$. The rotation from $(W_L^3, B)$ to $(Z_l, A)$ is the same as in the SM. We define the weak mixing angle as usual,
\begin{equation}
    \sin \theta_W = \frac{g_Y}{\sqrt{g^2+g_Y^2}} \, , \quad \cos \theta_W = \frac{g}{\sqrt{g^2+g_Y^2}} \, ,
\end{equation}
so that
\begin{align}
    &Z_l = \frac{1}{\sqrt{g^2+g_Y^2}}\left(g\hspace{0.3mm} W_L^3 - g_Y B\right)\;,&&
    A = \frac{1}{\sqrt{g^2+g_Y^2}}\left(g_Y W_L^3 +g\hspace{0.3mm} B\right)\;.
\end{align}
Applying the complete rotation from $(W^3_1,W^3_2,B)$ to $(Z_h,Z_l,A)$ results in the mass matrix
\begin{align} \label{eq:MnQuasidiag}
\arraycolsep=3pt\def\arraystretch{2.2}
    \mathcal{M}_n = \frac{1}{4}\left(\begin{array}{ccc}
        \frac{1}{4}v_\Omega^2(g_1^2+g_2^2) + v_H^2 \frac{g^2g_1^2}{g_2^2} & v_H^2 g_1^2 \sqrt{\frac{g^2+g_Y^2}{g_1^2+g_2^2}} & 0  \\
        v_H^2 g_1^2 \sqrt{\frac{g^2+g_Y^2}{g_1^2+g_2^2}} & v_H^2 (g^2+g_Y^2) & 0 \\
        0 & 0 & 0
    \end{array}\right)\;.
\end{align}
Finally, the mass eigenstates $(Z',Z)$ are given in terms of $(Z_h, Z_l)$ as
\begin{equation}
Z' = \cos \theta_n \, Z_h - \sin \theta_n \, Z_l \, , \quad Z = \sin \theta_n \, Z_h + \cos \theta_n \, Z_l \, .
\end{equation}
The mixing angle $\theta_n$ and the masses $M_Z$ and $M_{Z'}$ can be determined exactly by diagonalizing the mass matrix in Eq.~\eqref{eq:MnQuasidiag}. However, it is more illustrative to take advantage of the VEV hierarchy in Eq.~\eqref{eq:VEVhier} and obtain approximate expressions by expanding in powers of the small parameter
\begin{equation} \label{eq:epsomega}
\epsilon_\Omega = \frac{v_H}{v_\Omega} \, .
\end{equation}
One finds
\begin{equation} \label{eq:thetan}
 \theta_n = \frac{4 \, g^3}{g_1 \, g_2^3} \, \sqrt{g^2+g_Y^2} \, \epsilon_\Omega^2 + \mathcal{O}(\epsilon_\Omega^4) \, ,   
\end{equation}
as well as
\begin{align}
M_Z^2 &= \frac{1}{4}(g^2+g_Y^2) \left(1 - 4 \, \frac{g^4}{g_2^4} \, \epsilon_\Omega^2 \right) v_H^2 + \mathcal{O}(\epsilon_\Omega^4) \, , \\
M_Z'^2 &= \frac{1}{16}(g_1^2+g_2^2) \left(1 + 4 \, \frac{g^4}{g_2^4} \, \epsilon_\Omega^2 \right) v_\Omega^2 + \mathcal{O}(\epsilon_\Omega^4) \, ,
\end{align}
We note that in the limit $\epsilon_\Omega \to 0$ one recovers the SM expression for $M_Z$, as expected.

\subsubsection*{Charged gauge bosons}

We now consider $V_c = (W_1^+, W_2^+)$ and its complex conjugate $V_c^* = (W_1^-, W_2^-)$. In this basis, the charged gauge bosons mass matrix takes the form
\begin{align}
    \mathcal{M}_c = \frac{1}{16}\left(\begin{array}{cc}
        4 g_1^2v_H^2 + \frac{1}{2} g_1^2(v_\xi^2+v_\Omega^2 ) & - g_1g_2 v_\xi v_\Omega  \\
        - g_1g_2 v_\xi v_\Omega & \frac{1}{2}g_2^2(v_\xi^2+v_\Omega^2)
    \end{array}\right)\;.
\end{align}
Again, we want to rotate from $(W_1^+, W_2^+)$ to a basis $(W_h, W_l)$, where we can identify $W_l$ with the usual $SU(2)_L$ charged gauge boson. This can be achieved by setting $v_\Omega = v_\xi$ and $v_H = 0$, and then proceeding analogously as with the neutral gauge bosons. One finds
\begin{align}
    &W_h = \frac{1}{\sqrt{g_1^2+g_2^2}}\left(g_1 W_1^+-g_2W_2^+\right)\;,&&
    W_l = \frac{1}{\sqrt{g_1^2+g_2^2}}\left(g_2 W_1^+ +g_1W_2^+\right)\;.
\end{align}
In the basis $(W_h, W_l)$, the mass matrix is given by
\begin{align}
    \mathcal{M}_c = \frac{1}{16}\left(\begin{array}{cc}
        \frac{g^2g_1^2}{g_2^2}\left[4v_H^2+\frac{1}{2}\left(v_\xi+v_\Omega\frac{g_2^2}{g_1^2}\right)^2+\frac{1}{2}\left(v_\Omega+v_\xi\frac{g_2^2}{g_1^2}\right)^2\right] &\frac{g^2g_1}{g_2}\left[4v_H^2+\frac{1}{2}(v_\xi-v_\Omega)^2\left(1-\frac{g_2^2}{g_1^2}\right)\right] \\
        \frac{g^2g_1}{g_2}\left[4v_H^2+\frac{1}{2}(v_\xi-v_\Omega)^2\left(1-\frac{g_2^2}{g_1^2}\right)\right] &  \left[4v_H^2+(v_\xi-v_\Omega)^2\right]g^2
    \end{array}\right)\, . \label{eq:mcharged}
\end{align}
Finally, the charged mass eigenstates $(W',W)$ are given in terms of $(W_h, W_l)$ as
\begin{equation}
W' = \cos \theta_c \, W_h - \sin \theta_c \, W_l \, , \quad W = \sin \theta_c \, W_h + \cos \theta_c \, W_l \, .
\end{equation}
The mixing angle $\theta_c$ and the masses $M_W$ and $M_{W'}$ can be determined exactly by diagonalizing the matrix in Eq.~\eqref{eq:mcharged}. However, we again prefer to derive approximate expressions by expanding in powers of the small parameters $\epsilon_\Omega$, defined in Eq.~\eqref{eq:epsomega}, and
\begin{equation} \label{eq:epsdelta}
\epsilon_\Delta = \frac{v_\Omega - v_\xi}{v_\Omega} \, .
\end{equation}
The mixing angle is found to be
\begin{equation} \label{eq:thetac}
    \theta_c= -\frac{4 \, g^3}{g_1 \, g_2^3} \left[\epsilon_\Omega^2+\frac{1}{8} \epsilon_\Delta^2 \left(1-\frac{g_2^2}{g_1^2}\right)\right] + \mathcal{O}(\epsilon^4) \, ,
\end{equation}
where $\mathcal{O}(\epsilon^4)$ includes terms of order $\epsilon_\Omega^4$, $\epsilon_\Delta^4$ and $\epsilon_\Omega^2 \, \epsilon_\Delta^2$ or higher, whereas for the masses one finds
\begin{align}
M_W^2 &=\frac{1}{4} g^2 \left(1 + \frac{1}{4} \frac{\epsilon_\Delta^2}{\epsilon_\Omega^2} + \frac{g_1^2 (g_2^2-g_1^2)}{(g_1^2+g_2^2)^2} \, \epsilon_\Delta^2 \right) v_H^2 + \mathcal{O}(\epsilon^4) \, , \\
M_W'^2 &= \frac{1}{16}(g_1^2+g_2^2) \left[1-\epsilon_\Delta+4\frac{g^4}{g_2^4}\epsilon_\Omega^2+\frac{1}{2}\left(1-2\frac{g^4}{g_1^2g_2^2}\right)\epsilon_\Delta^2
\right] v_\Omega^2 + \mathcal{O}(\epsilon^4) \, .
\end{align}
We note that $M_W$ includes the ratio $\epsilon_\Delta^2/\epsilon_\Omega^2 = (v_\Omega-v_\xi)^2/v_H^2$, expected to be very small due to $v_\Delta \sim v_\Omega-v_\xi \ll v_H$ in the full theory, see Eq.~\eqref{eq:VEVhier}. Moreover, the SM expression for $M_W$ is recovered in the limit $\epsilon_\Delta \to 0$. 

\subsection{Scalar masses}
\label{subsec:scmass}

After SSB, the scalar fields split into two separate sets. On the one hand, the components with the same electric charge in the $H$ and $\Omega$ multiplets mix, since they are all even under the accidental \z2 symmetry of the model. On the other hand, the $\eta$ fields form an independent sector, since they are odd.

\subsubsection*{\z2-even scalar masses}

The neutral \z2-even scalar fields can be decomposed as
\begin{align}
\begin{aligned}
H^0 &= \frac{1}{\sqrt{2}} \left( v_H + S_H + i \, A_H \right) \, , \\
\Omega^0 &= \frac{1}{\sqrt{2}} \left( v_\Omega + S_\Omega + i \, A_\Omega \right)
\, , \\
\xi_0 &= \frac{1}{\sqrt{2}} \left( v_\Omega + S_\xi \right) \, .
\end{aligned}
\end{align}
We note that the self-duality condition imposed on $\Omega$ implies that $\xi_0$ is a real field. In the following, and just for the sake of simplicity, we will assume that CP is conserved in the scalar sector. Therefore, the CP-even and CP-odd states do not mix. In this case, one can define the bases
\begin{align}
\begin{aligned}
 \Ss^T & \equiv \left( S_H, S_\Omega, S_\xi \right) \quad , \qquad &\Pp^T& \equiv \left( A_H, A_\Omega \right) \, , \\[3mm]
 (\Hh^{-})^T & \equiv \left( H^- , \Omega_1^- , \Omega_2^- \right) \quad , \qquad &(\Hh^{+})^T& \equiv \left( H^+ , \Omega_1^+ , \Omega_2^+ \right) \, ,
\end{aligned}
\end{align}
and write the scalar mass Lagrangian
\begin{equation}
-\L_{m}^s = \frac{1}{2} \, \Ss^T \M_{\Ss}^2 \, \Ss + \frac{1}{2} \, \Pp^T \M_{\Pp}^2 \, \Pp + \left(\Hh^-\right)^T \M_{\Hh^\pm}^2 \Hh^+ + \Omega^{--} M_{\Omega^{++}}^2 \, \Omega^{++} \, .
\end{equation}
The mass matrix for the CP-even \z2-even scalars is given by
\begin{align}
    \M_{\Ss}^2 = \left(\begin{array}{ccc}
        \left(\M_{\Ss}^2\right)_{HH} & \left(\M_{\Ss}^2\right)_{H\Omega} & \left(\M_{\Ss}^2\right)_{H\xi}  \\
        \left(\M_{\Ss}^2\right)_{H\Omega} & \left(\M_{\Ss}^2\right)_{\Omega\Omega} & \left(\M_{\Ss}^2\right)_{\Omega\xi} \\
        \left(\M_{\Ss}^2\right)_{H\xi} & \left(\M_{\Ss}^2\right)_{\Omega\xi} & \left(\M_{\Ss}^2\right)_{\xi\xi}
    \end{array}\right)\;,
\end{align}
with
\begin{align}
    &\left(\M_{\Ss}^2\right)_{HH} = m_H^2 + \frac{3}{2}v_H^2\lambda_H + \frac{1}{64}(v_\xi^2+2v_\Omega^2)\lambda_{H\Omega} +\frac{3v_H^2v_\Omega^2\lambda_{H\Delta\Omega}^2}{64 m_\Delta^2} \;, \\[3mm]
    &\left(\M_{\Ss}^2\right)_{H\Omega} = \frac{v_Hv_\Omega}{4\sqrt{2}}\left(2\lambda_{H\Delta}+\frac{v_H^2\lambda_{H\Delta\Omega}^2}{m_\Delta^2}\right) \;, \\[3mm]
    &\left(\M_{\Ss}^2\right)_{H\xi} = \frac{v_Hv_\xi\lambda_{H\Omega}}{2\sqrt{2}}\;, \\[3mm]
    &\left(\M_{\Ss}^2\right)_{\Omega\Omega} = 2m_\Omega^2+ v_H^2\lambda_{H\Omega} + \frac{1}{32}(v_\xi^2+6v_\Omega^2)\lambda_{\Omega 1} + \frac{3}{32}v_\Omega^2\lambda_{\Omega 2} + \frac{2}{3}v_\xi\mu_1 +\frac{v_H^4\lambda_{H\Delta\Omega}^2}{4 m_\Delta^2}\;, \\[3mm]
    &\left(\M_{\Ss}^2\right)_{\Omega\xi} = \frac{1}{8}v_\Omega(v_\xi\lambda_{\Omega 1}+12\mu_1)\;, \\[3mm]
    &\left(\M_{\Ss}^2\right)_{\xi\xi} = 4m_\Omega^2+2v_H^2\lambda_{H\Omega}+\frac{1}{16}(3 v_\xi^2 + 2v_\Omega^2)\lambda_{\Omega 1} + \frac{3}{16}v_\xi^2\lambda_{\Omega 2}\;.
\end{align}
Since the tadpole equations~\eqref{eq:tadHEFT}-\eqref{eq:tadXiEFT} are linear in the parameters $m_H^2$, $m_\Omega^2$ and $1/m_\Delta^2$, one can easily solve for them. Replacing the solutions in the previous expressions for the elements of $\M_{\Ss}^2$ leads to
\begin{align}
    &\left(\M_{\Ss}^2\right)_{HH} = v_H^2\lambda_H+\frac{v_\Omega^2(v_\xi^2-v_\Omega^2)(v_\xi\lambda_{\Omega 2}-24\mu_1)}{256v_H^2 v_\xi} \;, \\[3mm]
    &\left(\M_{\Ss}^2\right)_{H\Omega} = \frac{1}{2\sqrt{2}}v_\Omega v_H\lambda_{H\Omega}+ \frac{v_\Omega(v_\xi^2-v_\Omega^2)(v_\xi\lambda_{\Omega 2}-24\mu_1)}{32\sqrt{2}v_H v_\xi} \;, \\[3mm]
    &\left(\M_{\Ss}^2\right)_{H\xi} = \frac{v_Hv_\xi\lambda_{H\Omega}}{2\sqrt{2}}\;, \\[3mm]
    &\left(\M_{\Ss}^2\right)_{\Omega\Omega} = \frac{1}{16}v_\Omega^2(2\lambda_{\Omega1}+\lambda_{\Omega2})\;, \\[3mm]
    &\left(\M_{\Ss}^2\right)_{\Omega\xi} = \frac{1}{8}v_\Omega(v_\xi\lambda_{\Omega 1}+12\mu_1)\;, \\[3mm]
    &\left(\M_{\Ss}^2\right)_{\xi\xi} = \frac{1}{8}v_\xi^2(\lambda_{ \Omega 1}+\lambda_{\Omega 2})-\frac{4}{3}\frac{v_\Omega^2\mu_1}{v_\xi}\;.
\end{align}
Obtaining approximate analytical expressions for the eigenvalues of this matrix would not be very enlightening. Instead, we note that terms proportional to the difference $v_\Omega^2-v_\xi^2$ are small, as already explained above. Therefore, if $\lambda_{H\Omega} \ll 1$ in order to suppress the mixing of $S_H$ with the rest of states, the lightest CP-even mass eigenstate $\Ss_1 \equiv h$ is a SM-like Higgs, which can be easily identified with the scalar with $m_h \approx 125$ GeV, discovered at the LHC. There are other two CP-even \z2-even scalars, $\Ss_2$ and $\Ss_3$, with masses of the order of $v_\Omega$. Let us now consider the the CP-odd \z2-even scalars. Similarly, in the Landau gauge, their mass matrix is given by
\begin{align}
    \M_{\Pp}^2 = \left(\begin{array}{cc}
        \left( \M_{\Pp}^2 \right)_{HH} & 0  \\
        0 & \left( \M_{\Pp}^2 \right)_{\Omega\Omega} \\
    \end{array}\right)\;,
\end{align}
with
\begin{align}
    &\left( \M_{\Pp}^2 \right)_{HH} = m_H^2 + \frac{1}{2}v_H^2\lambda_H+\frac{1}{64}(v_\xi^2+2v_\Omega^2)\lambda_{H\Omega}+\frac{v_H^2v_\Omega^2\lambda_{H\Delta\Omega}^2}{64 m_\Delta^2}\;,\\[3mm]
    &\left( \M_{\Pp}^2 \right)_{\Omega\Omega} = 2m_\Omega^2+v_H^2\lambda_{H\Omega}+\frac{1}{32}(v_\xi^2+2v_\Omega^2)\lambda_{\Omega 1}+\frac{1}{32}v_\Omega^2\lambda_{\Omega 2}+ \frac{2}{3} v_\xi\mu_1 + \frac{v_H^2\lambda_{H\Delta\Omega}^2}{4 m_\Delta^2}\;.
\end{align}
It is easy to check that replacing the solution of the tadpole equations in Eqs.~\eqref{eq:tadHEFT} and \eqref{eq:tadOmegaEFT} into $\M_{\Pp}^2$ leads to two vanishing eigenvalues. These are the Goldstone bosons that become the longitudinal components of the massive $Z$ and $Z'$ bosons. The mass matrix for the \z2-even singly charged scalars is given by
\begin{align}
    \M_{\Hh^\pm}^2 = \left(\begin{array}{ccc}
        \left(\M_{\Hh^\pm}^2\right)_{HH} & \left(\M_{\Hh^\pm}^2\right)_{H\Omega_1} & \left(\M_{\Hh^\pm}^2\right)_{H\Omega_2} \\
        \left(\M_{\Hh^\pm}^2\right)_{H\Omega_1} & \left(\M_{\Hh^\pm}^2\right)_{\Omega_1\Omega_1} & \left(\M_{\Hh^\pm}^2\right)_{\Omega_1\Omega_2} \\
        \left(\M_{\Hh^\pm}^2\right)_{H\Omega_2} & \left(\M_{\Hh^\pm}^2\right)_{\Omega_1\Omega_2} & \left(\M_{\Hh^\pm}^2\right)_{\Omega_2\Omega_2} 
        \end{array}\right)\;,
\end{align}
with
\begin{align}
    &\left(\M_{\Hh^\pm}^2\right)_{HH} = m_H^2+\frac{1}{2}v_H^2\lambda_H + \frac{1}{64}(v_\xi^2+2v_\Omega^2)\lambda_{H\Omega}+\frac{v_H^2v_\xi^2\lambda_{H\Delta\Omega}^2}{64 m_\Delta^2}\;,\\[3mm]
    &\left(\M_{\Hh^\pm}^2\right)_{H\Omega_1} = -\frac{v_H^3v_\xi\lambda_{H\Delta\Omega}^2}{16 m_\Delta^2}\;,\\[3mm]
    &\left(\M_{\Hh^\pm}^2\right)_{H\Omega_2} = \frac{v_H^3v_\Omega\lambda_{H\Delta\Omega}^2}{16 m_\Delta^2}\;,\\
    &\left(\M_{\Hh^\pm}^2\right)_{\Omega_1\Omega_1} = 2m_\Omega^2+v_H^2\lambda_{H\Omega}+\frac{1}{32}(v_\xi^2+2v_\Omega^2)\lambda_{\Omega 1}+\frac{1}{32}(v_\xi^2+v_\Omega^2)\lambda_{\Omega 2}+\frac{v_H^4\lambda_{H\Delta\Omega}^2}{4m_\Delta^2}\;,\\[3mm]
    &\left(\M_{\Hh^\pm}^2\right)_{\Omega_1\Omega_2} = -\frac{1}{32}v_\Omega(v_\xi\lambda_{\Omega 2}-24\mu_1)\;,\\[3mm]
    &\left(\M_{\Hh^\pm}^2\right)_{\Omega_2\Omega_2} = 2m_\Omega^2+v_H^2\lambda_{H\Omega}+\frac{1}{32}(v_\xi^2+2v_\Omega^2)\lambda_{\Omega 1}+\frac{1}{32}(v_\xi^2+v_\Omega^2)\lambda_{\Omega 2}\;.
\end{align}
Again, replacing Eqs.~\eqref{eq:tadHEFT}-\eqref{eq:tadXiEFT} into $\M_{\Hh^\pm}^2$ leads to two vanishing eigenvalues, in this case corresponding to the Goldstone bosons that become the longitudinal components of the $W$ and $W'$ bosons. We note, however, the existence of a physical charged scalar in the spectrum. We will simply denote it as $\Hh^+$. Its mass is given by
\begin{align}
 M_{\Hh^+}^2=& \, \frac{v_\xi\lambda_{\Omega 2}-24\mu_1}{512v_H^2 v_\xi} \, \left[16v_H^2(v_\xi^2+v_\Omega^2)+(v_\xi^2-v_\Omega^2)^2\right] \nonumber \\
 =& \frac{v_\xi\lambda_{\Omega 2}-24\mu_1}{512 v_\xi} \, \left[ 16 (v_\xi^2+v_\Omega^2)+ \frac{\epsilon_\Delta^2}{\epsilon_\Omega^2} (v_\xi+v_\Omega)^2 \right]\, .
\end{align}
Finally, the doubly charged scalar $\Omega^{++}$ has a mass
\begin{align}
    M_{\Omega^{++}}^2 =& \, 2m_\Omega^2 + v_H^2\lambda_{H\Omega}+\frac{1}{32}(v_\xi^2+2v_\Omega^2)\lambda_{\Omega 1} + \frac{3}{32}v_\Omega^2\lambda_{\Omega 2}-\frac{2}{3}v_\xi\mu_1+\frac{v_H^4\lambda_{H\Delta\Omega}^2}{4 m_\Delta^2} \nonumber \\
    =& \frac{1}{16}(v_\Omega^2\lambda_{\Omega2}-24v_\xi \mu_1) \;,
\end{align}
where in the last line we have applied the tadpole equations.

\subsubsection*{\z2-odd scalar masses}

The neutral $\eta_0$ field can also be decomposed into its real and imaginary components as
\begin{equation}
    \eta_0 = \frac{1}{\sqrt{2}} \left( \eta_R + i \, \eta_I \right) \, .
\end{equation}
Again, under the assumption of CP conservation in the scalar sector, these two components do not mix. In this case, their masses are found to be
\begin{align}
&m_R^2 = m_\eta^2 + \frac{1}{2}v_H^2\lambda_{H\eta} + \frac{1}{64}(v_\xi^2+2v_\Omega^2)\lambda_{\eta\Omega}+\frac{v_H^2v_\Omega\lambda_{H\Delta\Omega}\mu_2}{8m_\Delta^2}\;, \label{eq:mR}\\[3mm]
&m_I^2 = m_\eta^2 + \frac{1}{2}v_H^2\lambda_{H\eta} + \frac{1}{64}(v_\xi^2+2v_\Omega^2)\lambda_{\eta\Omega}-\frac{v_H^2v_\Omega\lambda_{H\Delta\Omega}\mu_2}{8m_\Delta^2}\;, \label{eq:mI}
\end{align}
Therefore, we find
\begin{equation}
m_R^2 - m_I^2 = \frac{v_H^2v_\Omega\lambda_{H\Delta\Omega}\mu_2}{4m_\Delta^2} = \lambda_5 \, v_H^2 \, ,
\end{equation}
where the effective $\lambda_5$ coupling was defined in Eq.~\eqref{eq:lam5}. Then, both mass eigenstates are degenerate in the limit $\lambda_5 \to 0$. Finally, as for the charged $\eta^+$ field, its mass is given by
\begin{align}
    m_+^2 = m_\eta^2 + \frac{1}{2}v_H^2\lambda_{H\eta} + \frac{1}{64}(v_\xi^2+2v_\Omega^2)\lambda_{\eta\Omega}\;.
\end{align}

\subsection{Fermion masses}
\label{subsec:fmass}

Quarks and charged leptons acquire masses as in the SM. Therefore, we will now concentrate on the \z2-odd fermions and the neutrinos.

\subsubsection*{\z2-odd fermion masses}

After $SU(2)_1\times SU(2)_2\to SU(2)_L$, the fermion bidoublet $\chi$ decomposes as $(\mathbf{2},\overline{\mathbf{2}})\to \mathbf{2}'\times\mathbf{2}' = \mathbf{3}'\oplus \mathbf{1}'$, where we use primes to denote $SU(2)_L$ representations. More explicitly:
\begin{equation}
    \chi = \left(\begin{array}{cc}
          \chi_0 & \chi_+ \\
          \chi_- & -\chi_0^c
    \end{array}\right) \xrightarrow{SU(2)_L} \left(\begin{array}{c}
        \chi_+  \\
        \frac{1}{\sqrt{2}}(\chi_0+\chi_0^c) \\
        \chi_-
    \end{array}\right) \oplus \frac{1}{\sqrt{2}}(\chi_0-\chi_0^c)
\end{equation}
The $\chi$ fermions receive two contributions to their mass: the bare mass term $M$ and the induced mass term generated by the $y_\chi$ Yukawa interaction after the $\Omega$ bitriplet gets a VEV. Their mass Lagrangian can be written as
\begin{equation}
    - \mathcal{L}_m^f = \frac{1}{2} \, \overline{\chi_n^c} \;\mathcal{M}_{\chi_n}\;\chi_n + \overline{\chi_c} \; \mathcal{M}_{\chi_c} \; \chi_c + \hc\;,
\end{equation}
where
\begin{equation}
    \chi_n = \left(\begin{array}{c}
         \chi_0  \\
         \chi_0^c
    \end{array}\right)\;,\quad \chi_c = \chi^+\;.
\end{equation}
The mass matrix for the neutral $\chi$ fermions is given by
\begin{equation}
\mathcal{M}_{\chi_n}=\left(\begin{array}{cc}
        -\frac{v_\Omega y_\chi }{8}& M +\frac{v_\xi y_\chi }{16}  \\
        M +\frac{v_\xi y_\chi }{16} & -\frac{v_\Omega y_\chi }{8} 
    \end{array}\right)\;.
\end{equation}
Diagonalizing $\mathcal{M}_{\chi_n}$ results in
\begin{equation}
    - \mathcal{L}_m^f \supset \frac{1}{2} \mathcal{M}_s\;\overline{\chi_s^c}\;\chi_s + \frac{1}{2} \mathcal{M}_t\;\overline{\chi_t^c}\;\chi_t\ + \hc \;, 
\end{equation}
where the states $\chi_s$ and $\chi_t$ are given by
\begin{align}
    \chi_s &= \frac{1}{\sqrt{2}}(\chi_0-\chi_0^c)\,, \\
    \chi_t &= \frac{1}{\sqrt{2}}(\chi_0+\chi_0^c)\,,
\end{align}
and their Majorana mass matrices are
\begin{align}
    \mathcal{M}_s &= M+\frac{1}{16}(2 v_\Omega + v_\xi) \, y_\chi = M + \frac{v_\Omega}{16} \, (3-\epsilon_\Delta) \, y_\chi \;, \\
    \mathcal{M}_t &= M-\frac{1}{16}(2 v_\Omega - v_\xi) \, y_\chi = M - \frac{v_\Omega}{16} \, (1+\epsilon_\Delta) \, y_\chi \;.
\end{align}
the mass matrix of the charged $\chi$ fermions is
\begin{equation}
\mathcal{M}_{\chi_c} = M - \frac{1}{16} \, v_\xi \, y_\chi = M - \frac{v_\Omega}{16} \, (1-\epsilon_\Delta) \, y_\chi \, .
\end{equation}
All $\chi$ fermions, neutral or charged, would be mass degenerate in the absence of the contribution induced by the $y_\chi$ Yukawa. This term introduces mass splittings, not only between neutral and charged $\chi$ fermions, but also within the neutral ones. We also note that $\mathcal{M}_{\chi_c} = \mathcal{M}_t$ in the $SU(2)_L$-conserving limit $\epsilon_\Delta \to 0$. This was expected, since $\chi_c$ and $\chi_t$ form an $SU(2)_L$ triplet. Finally, $\mathcal{M}_s$, $\mathcal{M}_t$ and $\mathcal{M}_{\chi_c}$ are $2\times 2$ matrices in generation space, which must be diagonalized to find the mass eigenstates. This can be achieved by means of four $2 \times 2$ unitary transformations, $U_s$, $U_t$, $U_L$ and $U_R$, defined by
\begin{align}
U_B^T \, \mathcal{M}_B \, U_B &= \widehat{\mathcal{M}}_B \, , \\
U_L^\dagger \, \mathcal{M}_{\chi_c} \, U_R &= \widehat{\mathcal{M}}_{\chi_c} \, ,
\end{align}
where $B=s,t$ and $\widehat{\mathcal{M}}_B$ and $\widehat{\mathcal{M}}_{\chi_c}$ denote the diagonal matrices in the mass basis. Therefore, the mass eigenstates are given by
\begin{align}
    \hat\chi_{B} &= U_{B} \, \chi_{B}\;, \\
    \left( \hat \chi_{\chi_c} \right)_L &= U_{L} \, \left( \chi_{\chi_c} \right)_L \;, \\
    \left( \hat \chi_{\chi_c} \right)_R &= U_{R} \, \left( \chi_{\chi_c} \right)_R \;.
\end{align}

\subsubsection*{Neutrino masses}

\begin{figure}[t!]
\centering
\includegraphics[width=1\textwidth]{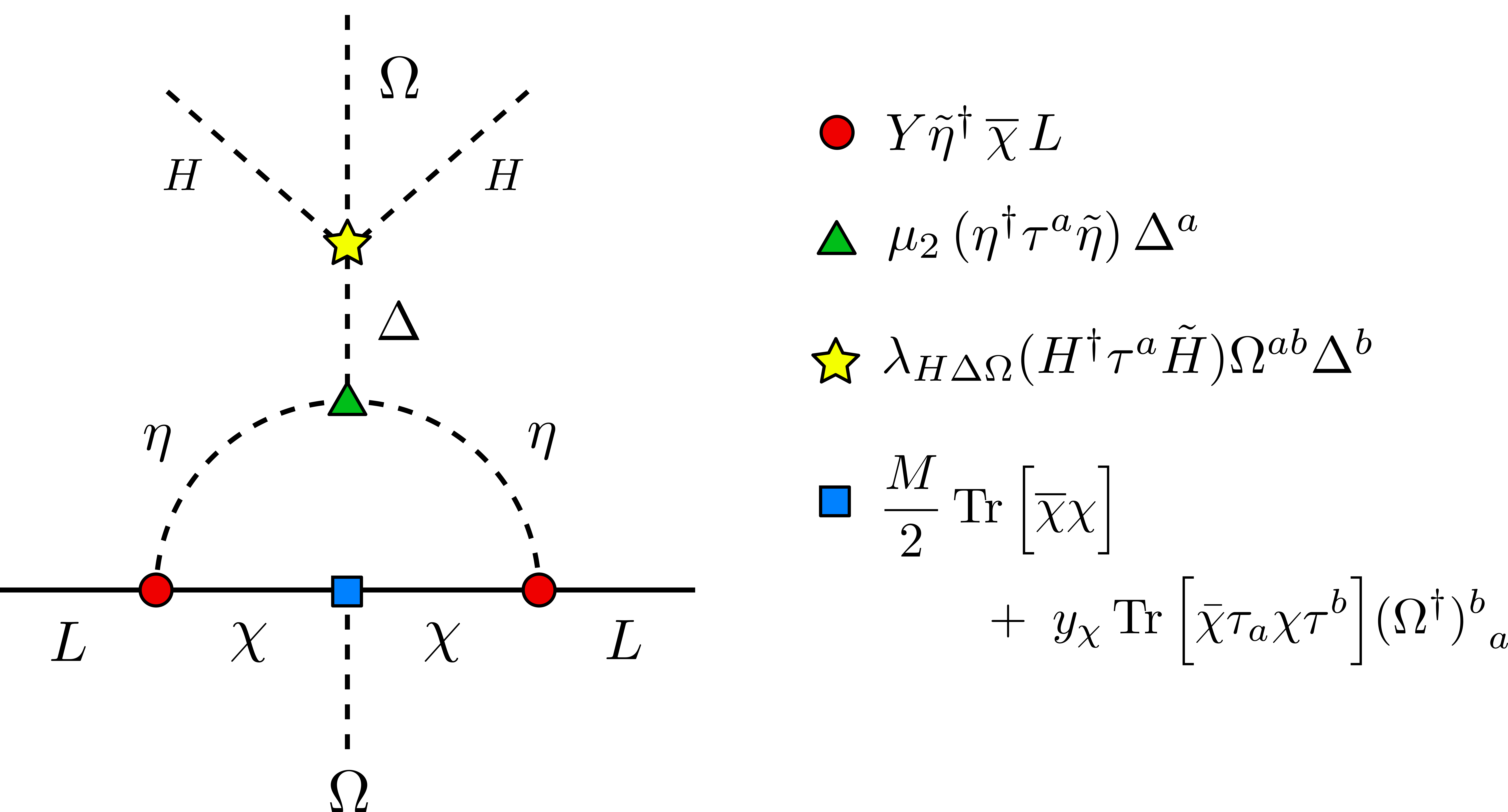}\vspace{5mm}
\caption{Neutrino mass generation in the full theory.
\label{fig:zanahoria}}
\end{figure}

\begin{figure}[t!]
\centering
\includegraphics[width=1\textwidth]{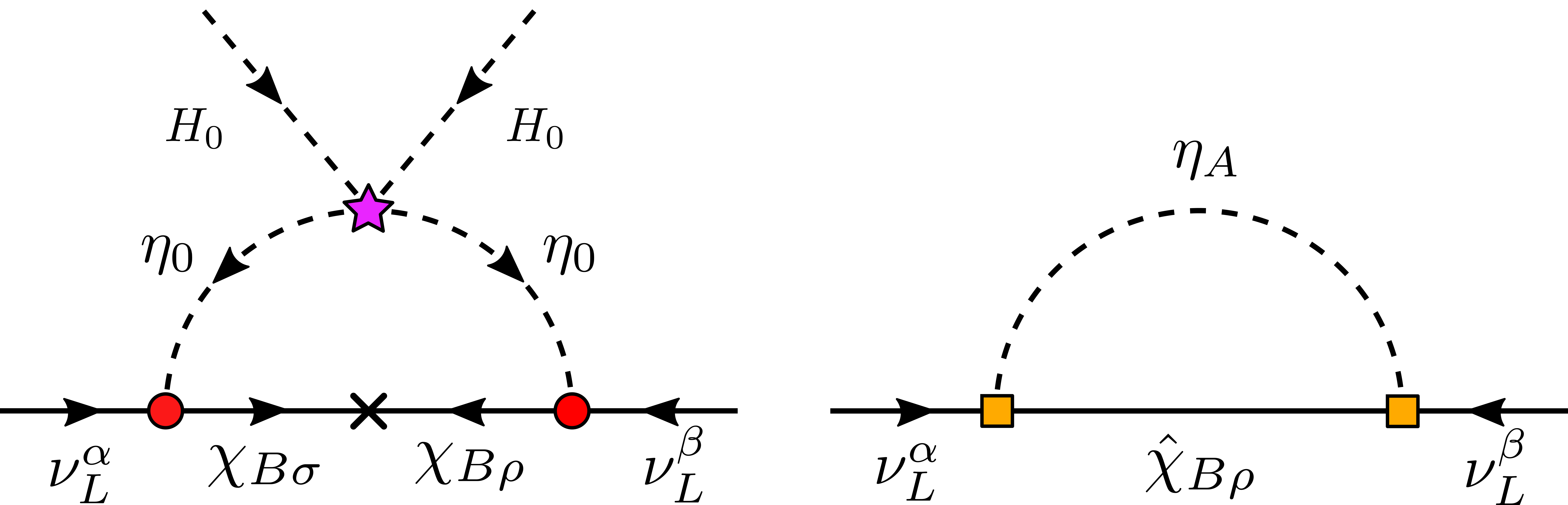}\vspace{5mm}
\caption{Neutrino mass generation in the low-energy theory. The only 1-loop contribution to neutrino masses, in the gauge (left) and mass bases (right). The star vertex in the gauge diagram is the effective $\lambda_5$ coupling, see Eq.~\eqref{eq:lam5}.
\label{fig:diagramas1loop}}
\end{figure}

In the full theory including $\Delta$, there exists a Weinberg-like operator shown in Fig.~\ref{fig:zanahoria}. After integrating out this heavy scalar triplet one finds a low-energy theory leading to the diagram shown in Fig.~\ref{fig:diagramas1loop}, which induces Majorana neutrino masses via the Scotogenic mechanism~\cite{Tao:1996vb,Ma:2006km}. The left-hand side of this figure displays gauge eigenstates, while the right-hand side contains the physical mass eigenstates. Charged loop diagrams do not contribute to the neutrino mass matrix. The neutrino mass matrix is given by the amplitude of the mass basis diagram summed over all indices inside the loop~\cite{Escribano:2020iqq},
\begin{equation} \label{eq:mnuini}
    -i\,m_\nu^{\alpha\beta} = \sum_{A,B,\rho}-i\,(m_\nu^{\alpha\beta})_{AB}^\rho =  \sum_{A,B,\rho}C_{\alpha \rho}^{AB}\int\frac{d^D k}{(2\pi)^D}\frac{i}{k^2-m_A^2}\frac{i(\slashed{k}+M_{B\rho})}{k^2-M_{B\rho}^2}C_{\beta \rho}^{AB}\;,
\end{equation}
where $A=R,I$, $B=s,t$ and $\rho=1,2$ and
\begin{itemize}
    \item $C_{\alpha \rho}^{AB}
    = \dfrac{i}{2} \, Y_{\alpha\sigma} \, \left(U_B\right)_{\sigma\rho} \, \kappa_A$, with $\kappa_A=1,i$ for $A=R,I$, is the coupling at the $\nu_\alpha-\eta_A-\hat{\chi}_{B \rho}$ vertex,
    \item $m_A^2 \equiv m_0^2 \pm \dfrac{1}{2}v_H^2\lambda_5$, where the $+$ ($-$) sign corresponds to $\eta_R$ ($\eta_I$), are given in Eqs.~\eqref{eq:mR} and \eqref{eq:mI},
    \item $M_{B\rho} = \left(\widehat{\mathcal{M}}_B \right)_{\rho \rho}$ is the mass of the $\hat \chi_{B\rho}$ neutral fermion.
\end{itemize}
Inserting everything into Eq.~\eqref{eq:mnuini} results in
\begin{align} 
    m_\nu^{\alpha\beta} =& \frac{1}{16\pi^2}\sum_{B,\rho,\sigma,\lambda} \left[ \frac{i}{2} Y_{\alpha\sigma} \left(U_B\right)_{\sigma\rho} \right] \left[ \frac{i}{2} Y_{\beta\lambda} \left(U_B \right)_{\lambda\rho}\right] M_{B\rho} \left[ B_0(0,m_R^2,M_{B\rho}^2)-B_0(0,m_I^2,M_{B \rho}^2)\right] \nonumber \\
    =& -\frac{1}{64\pi^2} \sum_{B,\rho,\sigma,\lambda} Y_{\alpha\sigma} Y_{\beta\lambda} \left(U_B\right)_{\sigma\rho} \left(U_B \right)_{\lambda\rho} \, M_{B\rho} \left[ B_0(0,m_R^2,M_{B\rho}^2)-B_0(0,m_I^2,M_{B \rho}^2)\right] \;, \label{eq:mnuint}
\end{align}
where $B_0$ is the standard Passarino-Veltman function~\cite{Passarino:1978jh}
\begin{equation}
    B_0(0,m_A^2,M_{B\rho}^2) =\Delta_\epsilon+1- \frac{m_A^2\log m_A^2 - M_{B\rho}^2\log M_{B\rho}^2}{m_A^2-M_{B\rho}^2} \, .
\end{equation}
We now expand the neutrino mass matrix for small $v_H^2\lambda_5$ up to first order. The zeroth order term is exactly zero because $m_R = m_I$ at this order. The first order expression turns out to be
\begin{align} 
    m_\nu^{\alpha\beta} &= \frac{\lambda_5 v_H^2}{64\pi^2} \sum_{B,\rho,\sigma,\lambda} Y_{\alpha\sigma} Y_{\beta\lambda} \left(U_B\right)_{\sigma\rho} \left(U_B \right)_{\lambda\rho} \, \frac{M_{B\rho}}{m_0^2-M_{B\rho}^2}\left[1- \frac{M_{B\rho}^2}{m_0^2-M_{B\rho}^2}\log\frac{m_0^2}{M_{B\rho}^2}\right] \nonumber \\
    &\equiv \left( Y^T \, \Lambda \, Y \right)_{\alpha \beta} \;, \label{eq:mnufinal}
\end{align}
where the $2 \times 2$ symmetric matrix $\Lambda$ is defined as
\begin{equation}
\Lambda_{\sigma \lambda} = \frac{\lambda_5 v_H^2}{64\pi^2} \sum_\rho \left[ \left(U_s\right)_{\sigma \rho} \left(U_s\right)_{\lambda \rho} M_{s \rho} \, F(m_0^2,M_{s \rho}^2)+\left(U_t\right)_{\sigma \rho} \left(U_t\right)_{\lambda \rho} M_{t \rho} \, F(m_0^2,M_{t \rho}^2)\right] \, ,
\end{equation}
and
\begin{equation}
F(x,y) = \frac{1}{x-y}\left[1- \frac{y}{x-y}\log\frac{x}{y}\right] \, .
\end{equation}
Finally, the neutrino mass matrix in Eq.~\eqref{eq:mnufinal} can be brought to diagonal form by means of a Takagi factorization,
\begin{equation}
U^T \, m_\nu \, U = \widehat{m}_\nu \, ,
\end{equation}
where $\widehat{m}_\nu$ is the diagonal matrix containing the physical active neutrino masses and $U$ is a unitary transformation. We will work in the basis in which the charged lepton mass matrix (and, thus, the $y$ Yukawa matrix) is diagonal. In this case, $U$ is the usual lepton mixing matrix measured in neutrino oscillation experiments.  

Some comments are in order:
\begin{itemize}
\item Neutrino masses are induced by a sum of contributions, including those from all neutral $\chi$ fermions. One may think that it is similar to the singlet-triplet Scotogenic model~\cite{Hirsch:2013ola,Rocha-Moran:2016enp}, since both $SU(2)_L$ singlets and triplets fermions contribute. However, the fermionic $SU(2)_L$ singlets and triplets in our model have a common $SU(2)_1 \times SU(2)_2$ origin, namely the $\chi$ bidoublets. For this reason, they share the same Yukawa interactions. In the case of a single generation of $\chi$ fermions one finds $\left( Y^T \, \Lambda \, Y \right)_{\alpha \beta} \propto Y_\alpha \, Y_\beta$, which in turn implies that the matrix in Eq.~\eqref{eq:mnufinal} would be of rank $1$, in conflict with oscillation data, that requires at least two non-vanishing neutrino masses. This is why we introduced two generations of $\chi$ bidoublets.
\item Since we have taken the minimal choice of two $\chi$ generations, only two active neutrinos acquire a non-zero mass. However, both normal hierarchy (NH) and inverted hierarchy (IH) can be accommodated.
\item Eq.~\eqref{eq:mnufinal} reproduces the correct size for neutrino masses with new states not too far from the TeV scale and sizable Yukawa couplings. This is thanks to the suppression introduced by $\lambda_5 \ll 1$. As already explained, this coupling is effectively generated after integrating out the heavy $\Delta$ triplet, see Eq.~\eqref{eq:lam5}. Therefore, it is naturally small if $m_\Delta$ is a large energy scale, as assumed throughout this paper. This not only suppresses Majorana neutrino masses, but also all lepton number violating effects. 
\end{itemize}

\subsection{Summary of the spectrum}
\label{subsec:sumspec}

Having already discussed in detail the particle spectrum of the model, let us summarize it.

\paragraph{Gauge bosons: SM + $\boldsymbol{Z', W'}$.} In addition to the usual SM gauge bosons, the model includes heavy $Z'$ and $W'$, associated to the breaking $SU(2)_1 \to SU(2)_2 \to SU(2)_L$, with masses of the order of $v_\Omega$.

\paragraph{Scalars: $\boldsymbol{\Ss_1, \Ss_2, \Ss_3, \Hh^+, \Omega^{++}, \eta_R, \eta_I, \eta^+}$.} There are two types of scalars in the model:
\begin{itemize}
\item \z2-even: There are three CP-even physical scalars in the model. The lightest of them is a SM-like Higgs boson, $\Ss_1 \equiv h$. The other two are heavy neutral scalars with masses of the order of $v_\Omega$. There are no physical \z2-even CP-odd scalars in the spectrum. Finally, there is a physical singly charged scalar, $\Hh^+$ and a doubly charged scalar, $\Omega^{++}$, both with masses of the order of $v_\Omega$.
\item \z2-odd: The model features a CP-even ($\eta_R$), a CP-odd ($\eta_I$) and a charged ($\eta^+$) \z2-odd scalar. This sector is the same as in the usual Scotogenic model.
\end{itemize}

\paragraph{Fermions: SM + $\boldsymbol{\chi_{s1}, \chi_ {s2}, \chi_{t1}, \chi_ {t2}, \chi_{c1}^+, \chi_ {c2}^+}$.} In addition to the usual charged fermions of the SM, the model has new \z2-odd fermions: four neutral ($\chi_{s1}$, $\chi_{s2}$, $\chi_{t1}$ and $\chi_{t2}$) and two charged ($\chi_{c1}^+$ and $\chi_{c2}^+$). Their masses receive contributions from both a bare mass term $M$ and a Yukawa contribution. These fermions participate in neutrino mass generation, giving rise to two massive neutrinos (while one remains massless).

\section{Phenomenology}
\label{sec:pheno}

In this Section we discuss some of the most relevant experimental signatures of our model. While the list will not be exhaustive, it will serve to illustrate some of the possible signals of our setup and to highlight the most relevant constraints that must be respected.

\subsection{Electroweak precision data}
\label{subsec:ew}

Precision electroweak measurements pose severe bounds in models that extend the electroweak gauge group, such as ours. Here we will concentrate on constraints associated to the masses and mixings of the electroweak gauge bosons. In particular, we will use the current limits on the $\rho$ parameter and gauge mixing in the neutral and charged sectors to derive bounds on the new scales in our model, namely the $SU(2)_1 \times SU(2)_2$ breaking scale and the high-energy mass parameter $m_\Delta$.

Let us consider the $\rho$ parameter. Again, it proves convenient to expand in powers of $\epsilon_\Omega$ and $\epsilon_\Delta$. At tree-level, one finds
\begin{equation} \label{eq:rho}
    \rho = \frac{M_W^2}{M_Z^2\cos{\theta_W^2}} = 1 + \frac{\epsilon_\Delta^2}{4 \, \epsilon_\Omega^2} + \frac{g^4 \, \epsilon_\Delta^2}{g_1^2 \, g_2^2} + \frac{4 \, g^4 \, \epsilon_\Omega^2}{g_2^4} + \mathcal{O}(\epsilon^4) \, .
\end{equation}
This must be compared to the current determination of the $\rho$ parameter from global fits to all available electroweak data~\cite{ParticleDataGroup:2024cfk},
\begin{equation}
\rho = 1.00031 \pm 0.00019 \, ,
\end{equation}
which implies that deviations from $\rho=1$ can be at most of order $\sim 10^{-4}$. Imposing that all contributions in Eq.~\eqref{eq:rho} respect this limit and assuming $g_1 = g_2$ leads to
\begin{equation} \label{eq:boundEps}
\epsilon_\Omega \lesssim 10^{-2} \, , \quad \epsilon_\Delta \lesssim 2 \cdot 10^{-4} \, .
\end{equation}
These bounds can be translated into bounds on the relevant energy scales. First, $\epsilon_\Omega \lesssim 10^{-2}$ implies
\begin{equation}
v_\Omega \gtrsim 20 \, \text{TeV} \, .
\end{equation}
This limit on the $SU(2)_1 \times SU(2)_2$ breaking scale must be understood as a very conservative limit. In fact, the $\epsilon_\Omega$ contributions to the $\rho$ parameter can be reduced by taking $g_2 \gg g$, as can be seen in Eq.~\eqref{eq:rho}. We estimate that $v_\Omega$ can be reduced to $\sim 1-2$ TeV while keeping $g_2$ in the perturbative regime ($< \sqrt{4 \pi}$). One can also use the bound on $\epsilon_\Delta$ in Eq.~\eqref{eq:boundEps} to find a limit on the VEV difference $v_\Omega - v_\xi$ and, subsequently, on $m_\Delta$. This can be achieved by taking advantage of Eqs.~\eqref{eq:vDelta} and \eqref{eq:oomDelta}. Barring cancellations in Eq.~\eqref{eq:vDelta}, assuming that all dimensionful parameters are of order $v_\Omega$ (except, of course, $v_H$) and taking $\lambda_{H \Delta \Omega} \sim 1$, we find
\begin{equation}
v_\Omega - v_\xi \lesssim 5 \, \text{GeV} \, .
\end{equation}
On the other hand, the resulting lower bound on $m_\Delta$ depends on the value of $v_\Omega$ and, for some choices, it is not consistent with our assumption $v_\Omega,v_\xi \ll m_\Delta$. Therefore, in the following we will simply consider $m_\Delta$ scales well above $20$ TeV.

Our model also leads to mass mixing in the gauge sector. The $Z$ boson does not correspond exactly to the pure $SU(2)_L$ gauge boson $Z_l$, but has a non-zero component in $Z_h$. Similarly, the $W$ boson is an admixture of $W_l$ (the pure $SU(2)_L$ boson) and $W_h$. As a result of this, the $Z$ and $W$ couplings deviate from their SM values. Therefore, it is easy to guess that it is strongly constrained by experimental data. Indeed, the corrections to the gauge boson couplings are bounded by many experiments, including electroweak precision data collected at the $Z$- and $W$-pole at LEP as well as a plethora of flavor measurements. For a detailed discussion on this effect, usually referred to as \textit{gauge mixing}, we refer to~\cite{Boucenna:2016wpr}.

Gauge mixing is controlled in our model by the $\theta_n$ and $\theta_c$ angles, given in Eqs.~\eqref{eq:thetan} and \eqref{eq:thetac}, respectively. In order to derive the corresponding constraints on the parameters of our model, one would need to perform a complete analysis of all the observables affected by gauge mixing. This type of comprehensive analysis has been done before for models based on the $SU(2)_1\times SU(2)_2\times U(1)_Y$ electroweak symmetry~\cite{Boucenna:2016qad,Hue:2016nya}, finding
\begin{equation}
\theta_n , \theta_c \lesssim 10^{-3} \, .
\end{equation}
Since the resulting constraints on the mass $\epsilon_\Omega$ and $\epsilon_\Delta$ expansion parameters, or those on the energy scales of the model, are slightly less stringent than those derived from the $\rho$ parameter, we will not discuss them further.

\subsection{Lepton flavor violating decays}
\label{subsec:lfv}

\begin{figure}[t!]
\centering
\includegraphics[width=1\textwidth]{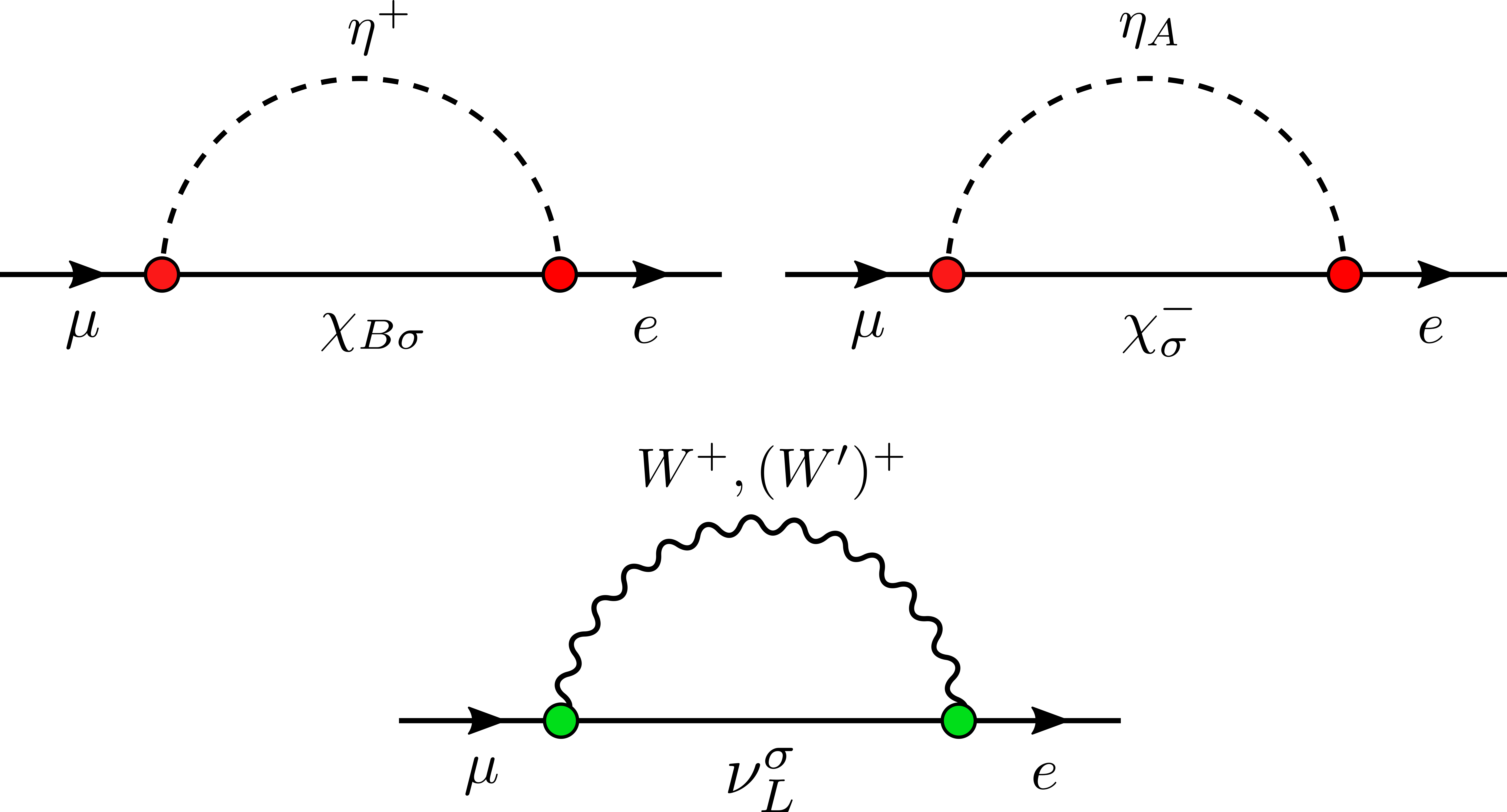}
\caption{1-loop contributions to the $\mu \to e \gamma$ rate in our model. The photon line is not drawn, but can be attached to all internal charged lines.
\label{fig:meg}}
\end{figure}

Since the new energy scales in our model are in the multi-TeV range, direct production of some of the heavy states is very unlikely, at least in the near future. This motivates searching for indirect signals of new physics in flavor observables.

In particular, we consider charged lepton flavor violating decays and, more precisely, the radiative channel $\mu \to e \gamma$. This is the most constraining LFV observable in most extensions of the lepton sector, since the new physics states typically induce large contributions to the $\bar e \sigma_{\mu \nu} e \, F^{\mu \nu}$ dipole operator. In our model, these are generated by the diagrams in Fig.~\ref{fig:meg}. The two diagrams with charged gauge bosons in the loop can be safely neglected, since their contributions to the $\mu \to e \gamma$ rate are suppressed by the usual $(m_\nu/m_W)^4$ factor. In our model, the light neutrinos do not mix with any other state. This, together with a unitary leptonic mixing matrix $U$, leads to this GIM suppression. We also note that neutral gauge bosons do not contribute at the 1-loop level, since they do not have flavor violating couplings (at tree-level). Therefore, we can concentrate on the two diagrams with scalars in the loop. Analytical expressions for their amplitudes can be found in~\cite{Lavoura:2003xp} and we have explicitly checked that they match those found in~\cite{Toma:2013zsa,Rocha-Moran:2016enp} in other Scotogenic models.

In the following, we explore the parameter space of our model and obtain some numerical results for BR$(\mu \to e \gamma)$. The current limit on this observable has been recently set by the MEG II experiment using data from their 2021-2022 physics runs. At 90\% C.L., the bound is given by~\cite{MEGII:2025gzr}
\begin{equation} \label{eq:megbound}
\text{BR}(\mu \to e \gamma) < 1.5 \cdot 10^{-13} \, .
\end{equation}
The MEG II collaboration plans to improve this limit with the inclusion of more data and reach a sensitivity to $\mu \to e \gamma$ branching ratios as low as $\sim 6 \cdot 10^{-14}$.

In our numerical analysis we fix the parameters
\begin{align}
y_\chi = 0.05 \, \id_2 \, , \quad \lambda_{H \eta} = 0.1 \, , \quad \lambda_{\eta \Omega} = 0.01 \, , \quad v_\Omega - v_\xi = 4 \, \text{GeV} \, ,
\end{align}
where $\id_n$ denotes the $n \times n$ identity matrix. In order to ensure that all points lead to correct neutrino masses and lepton mixing, we write the Yukawa matrix $Y$ in terms of neutrino oscillation parameters by means of a Casas-Ibarra parametrization~\cite{Casas:2001sr}, properly adapted to the specific version of the Scotogenic mechanism in our model as~\cite{Toma:2013zsa,Cordero-Carrion:2018xre,Cordero-Carrion:2019qtu}
\begin{equation} \label{eq:ci}
Y = V^\dagger \, \widehat{\Lambda}^{-1/2} \, R \, \left( \begin{array}{ccc} 0 & \sqrt{m_2} & 0 \\ 0 & 0 & \sqrt{m_3} \end{array} \right) \, P \, U^\dagger \, .
\end{equation}
Here $\Lambda = V^T \, \widehat{\Lambda} \, V$, where $V$ is a unitary matrix and $\widehat \Lambda$ is a diagonal matrix,~\footnote{We note that this Takagi factorization is always possible since $\Lambda$ is a symmetric matrix.}
\begin{equation} \label{eq:Pdef}
P = \left\{ \begin{array}{cl}
\id_3 \quad & \text{for NH} \, , \\
P_{13} \quad & \text{for IH} \, ,
\end{array} \right.
\end{equation}
with
\begin{equation} \label{eq:P13}
P_{13} = \begin{pmatrix} 0 & 0 & 1 \\ 0 & 1 & 0 \\ 1 & 0 & 0 \end{pmatrix} \, ,
\end{equation}
a permutation matrix. $m_2$ and $m_3$ are the active neutrino masses. In case of IH one should also rename $m_3 \to m_1$. Finally, $R$ is a $2 \times 2$ complex orthogonal matrix which can be parametrized by one complex angle. For the sake of simplicity, we will just take $R = \id_2$, focus on neutrino NH and use the best-fit values for the oscillation parameters determined by the global fit~\cite{deSalas:2020pgw}.

\begin{figure}[t!]
    \centering
    \includegraphics[width=0.65\linewidth]{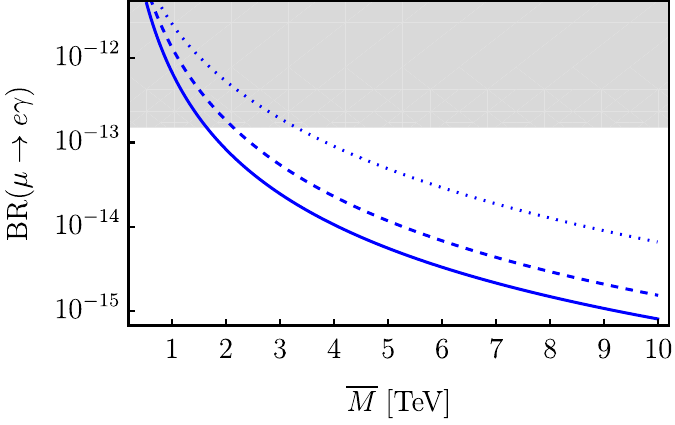}
    \caption{BR$(\mu \to e \gamma)$ as a function of $\overline M$ for three values of $m_\eta^2$: (200 GeV)$^2$ (blue), (1 TeV)$^2$ (blue, dashed) and (2 TeV)$^2$ (blue, dotted). The region in gray is excluded by the MEG II bound on BR$(\mu \to e \gamma)$, see Eq.~\eqref{eq:megbound}.
    \label{fig:M}}
\end{figure}  

We first study the dependence of BR$(\mu \to e \gamma)$ on the $\chi$ fermions' masses. To this end, we fix $v_\Omega = 20$ TeV and $\lambda_5 = 10^{-8}$ and define $M = \overline M \, \id_2$. Fig.~\ref{fig:M} shows BR$(\mu \to e \gamma)$ as a function of $\overline M$ for three values of $m_\eta^2$: (200 GeV)$^2$ (1 TeV)$^2$ and (2 TeV)$^2$. The gray region is experimentally excluded by MEG II, as it predicts a $\mu \to e \gamma$ branching ratio larger than that in Eq.~\eqref{eq:megbound}. As expected, the LFV rate decreases for higher $\overline M$ values, which imply higher $\chi$ masses, both for the neutral and charged fermions. Therefore, there is a lower bound on $\overline M$, with the precise value depending on the choice of $m_\eta^2$. Interestingly, larger $m_\eta^2$ does not necessarily lead to smaller BR$(\mu \to e \gamma)$. This is because the neutrino mass matrix in Eq.~\eqref{eq:mnufinal} links the values of the $Y$ Yukawa couplings to the $\chi$ and $\eta$ masses. Larger $m_\eta^2$ values actually imply larger $Y$ Yukawa parameters in order to keep the correct size of neutrino masses. This feature, enforced in our numerical scan by the Casas-Ibarra parametrization in Eq.~\eqref{eq:ci}, is behind the increase of BR$(\mu \to e \gamma)$ for larger $m_\eta^2$.

\begin{figure}[t!]
    \centering
    \includegraphics[width=0.5\linewidth]{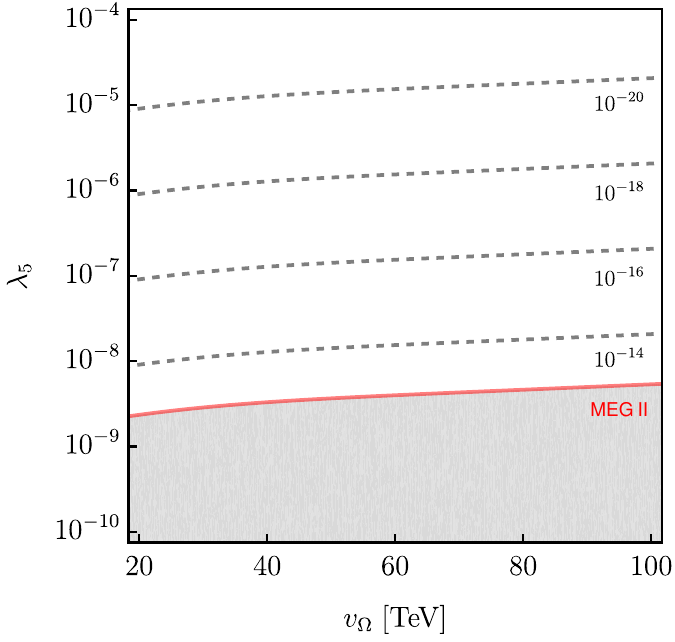}
    \includegraphics[width=0.48\linewidth]{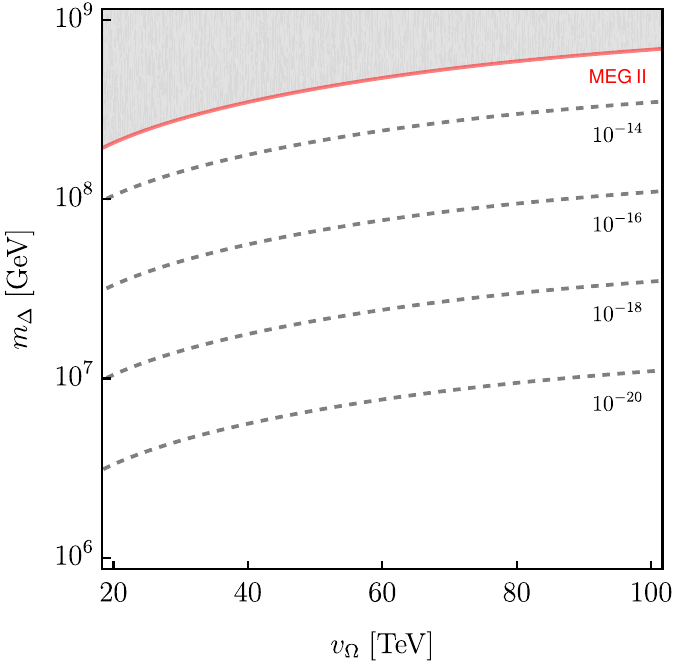}
    \caption{Contours of BR$(\mu \to e \gamma)$ in the $v_\Omega-\lambda_5$ (left) and $v_\Omega-m_\Delta$ (right) planes. The red line corresponds to the current MEG II limit on BR$(\mu \to e \gamma)$ given in Eq.~\eqref{eq:megbound}. The regions in gray are excluded by this bound.
    \label{fig:contours}}
\end{figure}

Let us turn our attention to the high energy scales in our model. Fig.~\ref{fig:contours} shows contours of BR$(\mu \to e \gamma)$ in the $v_\Omega-\lambda_5$ (on the left) and $v_\Omega-m_\Delta$ (on the right) planes. This figure has been obtained by fixing $M_{11} = 1$ TeV and $M_{22} = 1.5$ TeV, which leads to $\chi$ fermion masses in the TeV ballpark, and $m_\eta^2 = (500 \, \text{GeV})^2$. Both plots display the current limit on BR$(\mu \to e \gamma)$ given in Eq.~\eqref{eq:megbound} and include additional contour for some illustrative values of the observable. On the left panel we see that $\lambda_5$ values below $\sim 3 \cdot 10^{-9}$ are excluded, with a mild dependence on $v_\Omega$. The strong impact of the $\lambda_5$ parameter can again be understood by inspecting the neutrino mass matrix in Eq.~\eqref{eq:mnufinal}. Smaller $\lambda_5$ values imply larger $Y$ Yukawas to keep neutrino masses in agreement with data. This is common to the minimal Scotogenic model. In our model, $\lambda_5$ is an effective parameter, obtained after integrating out the heavy $\Delta$ triplet. Therefore, we can relate the size of $\lambda_5$ to the $m_\Delta$ mass scale, see Eq.~\eqref{eq:lam5}. Assuming $\lambda_{H \Delta \Omega} \sim 1$ and $\mu_2 \sim v_\Omega$ one finds $m_\Delta^2 \sim v_\Omega^2 / (4 \lambda_5)$. Therefore, a lower limit on $\lambda_5$ translates into an upper limit on $m_\Delta$, of about $\sim 2 \cdot 10^8$ GeV, as shown on the right panel of Fig.~\ref{fig:contours}. These results are generic in our model, although the exact numerical value for the limit on $m_\Delta$ depends on parameters such as $\lambda_{H \Delta \Omega}$ and $\mu_2$.

\subsection{Dark matter}
\label{subsec:dm}

The accidental \z2 of our model is an exact symmetry at all energies. Therefore, it can be used to stabilize a potentially valid DM candidate. Indeed, the lightest state that is charged under the \z2 is completely stable. We emphasize again that this is a direct consequence of the $SU(2)_1 \times SU(2)_2$ gauge symmetry and the choice of representations that we made. In fact, our model would be an example of \textit{predestined dark matter}~\cite{Ma:2018bjw}, a scenario that stabilizes a DM candidate by means of the gauge symmetry, without any additional imposed symmetry.

There are two scenarios:
\paragraph{Scalar DM.} $\eta_R$ or $\eta_I$, depending on the sign of the effective $\lambda_5$ coupling. Since these scalars belong to an $SU(2)_L$ doublet, they have gauge interactions with the light $Z$ and $W$ bosons, which allows them to be thermally produced in the early Universe. This scenario resembles the popular inert doublet model~\cite{Deshpande:1977rw}. Unless one takes advantage of specific resonant regions, the DM mass required to reproduce the observed relic density is found to be $\sim 500$ GeV, see for instance~\cite{Belyaev:2016lok}.

\paragraph{Fermion DM.} $\chi_{s1}$ or $\chi_{t1}$. In this case, the DM phenomenology may seem similar to that of the singlet-triplet Scotogenic model, recently studied in~\cite{Karan:2023adm}. However, there is a key difference. $\chi_{s1}$ and $\chi_{t1}$ are mass eigenstates and do not mix, even after electroweak symmetry breaking. Therefore, in the absence of gauge mixing (which is nevertheless constrained to be small, see Sec.~\ref{subsec:ew}), $\chi_{s1}$ does not couple to the light gauge bosons, while $\chi_{t1}$ couples as the neutral component of a pure triplet. As a consequence of this, the fermion DM scenario of our model leads either to a singlet fermion DM, as in the standard Scotogenic model, or to a triplet fermion DM, as in the variant of~\cite{Ma:2008cu}. The former case has been extensively studied, see~\cite{Avila:2025qsc} and references therein. The latter case is known to be very constrained, with the mass of the triplet DM candidate close to $\sim 2.3$ TeV in order to reproduce the observed relic density~\cite{Ma:2008cu}. Finally, the heavy $Z'$ and $W'$ gauge bosons may have an impact on the DM phenomenology if they are not too heavy, not far from the TeV scale, since both $\chi_{s1}$ and $\chi_{t1}$ couple to them.
\section{Summary and conclusion}
\label{sec:conclu}

We have presented a model based on the extended electroweak gauge group $SU(2)_1 \times SU(2)_2 \times U(1)_Y$, which naturally gives rise to the Scotogenic mechanism after symmetry breaking. As a result, our framework generates radiative Majorana neutrino masses and includes a viable dark matter candidate. In addition, it addresses several limitations of the conventional Scotogenic model. In particular, it explains the smallness of lepton number violation by linking it to the presence of heavy states, and it provides an origin for the \z2 parity, which emerges as an accidental symmetry of the theory.

Precision electroweak data, in particular the $\rho$ parameter, require the new scales in our model to be above the TeV. This makes the direct production of the new states at colliders unlikely. However, we have shown that one can probe the model indirectly with searches for lepton flavor violating decays, such as the radiative $\mu \to e \gamma$. In fact, the current MEG II limit already poses some limitations in the parameter space of the model. We also expect a rich dark matter phenomenology, especially in the case of a fermionic candidate.

There are several promising directions in which our model can be further investigated. Our analysis indicates that the $SU(2)_1 \times SU(2)_2$ breaking scale must be relatively high to satisfy the constraint from the $\rho$ parameter. However, in certain regions of the viable parameter space, this scale could lie only slightly above the TeV range. In such cases, one may expect a non-negligible production of $Z'$ and $W'$ bosons at the LHC. Additionally, exotic signatures arising from the production and decay of the doubly charged scalar $\Omega^{++}$, typically involving multiple charged leptons in the final state, could offer complementary probes. Finally, we also believe that the dark matter phenomenology of our model deserves further study, in particular to assess whether the new ingredients of our setup, such as the heavy gauge bosons, introduce qualitative differences with respect to simpler Scotogenic models.

\section*{Acknowledgements}

Work supported by the Spanish grants PID2023-147306NB-I00, CNS2024-154524
and CEX2023-001292-S (MICIU/AEI/10.13039/501100011033), as well as from CIPROM/2021/054 (Generalitat Valenciana). JPS is supported by the grant CIACIF/2022/158, funded by Generalitat Valenciana.

\appendix

\section{$\boldsymbol{SU(2)}$, real representations and the spherical basis}
\label{app:SU2stuff}

All $SU(2)$ representations~\footnote{There is an unfortunate custom of referring, individually, to both the matrices and the multiplets they act on as representations. Unless explicitly stated, representation will always mean the multiplets here.} with integer isospin $j=1,2,3,4\ldots$ are real representations. This means we can find a basis where the complex conjugate of a multiplet equals itself. Rotating them both by the same transformation $V$ would maintain this equivalence:
\begin{equation}
    \psi_r = \psi_r^* \;\xrightarrow{\quad}\; V\psi_r = 
    (V \psi_r)^* \; \xrightarrow{\quad}\; e^{i\alpha_a T_r^a}\psi_r = e^{-i\alpha_a(T_r^a)^*} \psi_r^* \;\xrightarrow{\psi_r^*=\psi_r}\;  (T_r^a)^* = -T_r^a\;,
\end{equation}
meaning the generators acting on the real representation are purely imaginary. For the $j=1$ triplet representation of $SU(2)$, a suitable choice of imaginary generators is
\begin{equation}
    T_r^1 = i\left(\begin{array}{ccc}
        0 & 0 & 0 \\
        0 & 0 & -1 \\ 
        0 & 1 & 0
    \end{array}\right)\;, \qquad T_r^2 = i\left(\begin{array}{ccc}
        0 & 0 & 1 \\
        0 & 0 & 0 \\ 
        -1 & 0 & 0
    \end{array}\right)\;, \qquad
    T_r^3 = i\left(\begin{array}{ccc}
        0 & -1 & 0 \\
        1 & 0 & 0 \\ 
        0 & 0 & 0
    \end{array}\right)\;,
\end{equation}
which are just the Hermitian version of the $SO(3)$ generators with all indices (both matrix and the $T^a$ triplet indices)\footnote{The set of all $SU(N)$ generators form a representation of the group themselves; the adjoint of dimensionality $2N+1$. For $SU(2)$ this means the three generators form a triplet, and so they get a triplet index $a$ on top.} in the cartesian basis. From now on, we will call this set of generators ``$SO(3)$ basis''. These generators are not the most useful when doing high-energy physics however, where $SU(2)$ multiplets are always written in a basis with a diagonal $T^3$ generator so that the fields embedded inside have a definite $T^3$ value, and thus definite electromagnetic charge via a Gell-Mann-Nishijima relation like $Q=T^3+Y$. To build real representations in our choice of $T_3$-diagonal basis, we have to find the change of basis matrix from the $SO(3)$ basis. The following (unitary) matrix diagonalises $T_r^3$ and is built from its eigenvectors in columns, which we choose to be
\begin{equation}\label{eq:sphericalTransformation}
    U = \left(\begin{array}{ccc}
        \mathbf{\hat{e}_1} & \mathbf{\hat{e}_0} & \mathbf{\hat{e}_{-1}} 
    \end{array}\right) = \frac{1}{\sqrt{2}}\left(\begin{array}{ccc}
        -1 & 0 & 1 \\
        -i & 0 & -i \\
        0 & \sqrt{2} & 0
    \end{array}\right)\;.
\end{equation}
The generators in the $T^3$-diagonal basis can be obtained by performing the similarity transformation $T^a = U^\dagger T_r^a U$:

\begin{align}
&T^1 = \frac{1}{\sqrt{2}}\left(\begin{array}{ccc}
        0 & 1 & 0 \\
        1 & 0 & 1 \\
        0 & 1 & 0
    \end{array}\right)\;,&& T^2 = \frac{1}{\sqrt{2}}\left(\begin{array}{ccc}
        0 & -i & 0 \\
        i & 0 & -i \\
        0 & i & 0
    \end{array}\right)\;,&& T^3 = \left(\begin{array}{ccc}
        1 & 0 & 0 \\
        0 & 0 & 0 \\
        0 & 0 & -1
    \end{array}\right)\;.
\end{align}
In a physics context, this set of generators is commonly said to be in the ``cartesian basis'', and we will refer to it by this name throughout this Appendix and the rest of the paper. This naming is due to the triplet index $T^a$ of the generators, which is still a cartesian index (it did not get transformed from the $SO(3)$ generators), and has nothing to do with the $ij$ matrix indices in $(T)_{ij}$, which in our new basis talk about the $T^3$ value of the triplet's components it acts on; $i,j=1,0,-1$. The basis spanned by the eigenvectors in Eq.\ (\ref{eq:sphericalTransformation}) is called the \textit{spherical basis}. It is related to the cartesian vector basis via the $U$ matrix; from Eq.\ (\ref{eq:sphericalTransformation}), it is easy to see that
\begin{equation}
    \left(\begin{array}{c}
         \mathbf{\hat{e}_1}  \\
         \mathbf{\hat{e}_0} \\
         \mathbf{\hat{e}_{-1}}
    \end{array}\right) = \frac{1}{\sqrt{2}}\left(\begin{array}{ccc}
        -1 & -i & 0 \\
        0 & 0 & \sqrt{2} \\
        1 & -i & 0
    \end{array}\right)\left(\begin{array}{c}
         \mathbf{e_1}  \\
         \mathbf{e_2} \\
         \mathbf{e_3}
    \end{array}\right) \;\xrightarrow{\text{ kets }}\; |\mathbf{\hat{e}}\rangle = U^T |\mathbf{e}\rangle\;,
\end{equation}
where we made a switch to bra-ket notation. The bra of a ket is just its Hermitian conjugate, so for some vector $\mathbf{A}$ we have
\begin{equation}\label{eq:sphericalComponentsTransformation}
    \mathbf{A} = \langle \mathbf{e}| A\rangle =  \langle\mathbf{\hat{e}}|U^T|A\rangle = \langle\mathbf{\hat{e}}|\hat{A}\rangle \;\implies\; |\hat{A}\rangle = U^T |A\rangle \;,
\end{equation}
meaning the components of a vector (in a column) transform via $U^T$ too. Now that we know how vector components transform, let us get back to the matter of real representations. We already discussed that in the original $SO(3)$ basis, multiplets are real in the sense of $\psi_r = \psi_r^*$. We are now working in the cartesian basis, and since multiplets are kets in their representation space, they will transform from the $SO(3)$ basis via $U^T$. For a real multiplet this would mean
\begin{equation}
    \psi_r = \psi_r^* \;\xrightarrow{\text{ cartesian }}\; (U^*\psi) = (U^*\psi)^* \;\xrightarrow{\text{ unitarity }}\; \psi = U^TU\psi^* \;\xrightarrow{\text{ define }}\; \psi = C\psi^* = \Tilde{\psi}\;,
\end{equation}
where in the last step we defined the $SU(2)$ \textit{conjugation matrix} $C=U^TU$ and the $SU(2)$ \textit{dual} $\Tilde{\psi}$. For the doublet\footnote{Although the $SU(2)$ doublet generators (the halved Pauli matrices) cannot be rotated into a purely imaginary set of generators, the dual of a doublet still exists. It cannot, however, be a real representation, since there is no way of satisfying $\psi=\tilde{\psi}$ with two real degrees of freedom.} and triplet representations, the conjugation matrix are
\begin{equation}
    C_2 = \left(\begin{array}{cc}
        0 & 1 \\
        -1 & 0 
    \end{array}\right)\;,\qquad
    C_3 = \left(\begin{array}{ccc}
        0 & 0 & -1 \\
        0 & 1 & 0 \\
        -1 & 0 & 0 \\
    \end{array}\right)\;.
\end{equation}
In summary, when working in a $T_3$-diagonal basis, the reality condition $\psi_r=\psi_r^*$ in the $SO(3)$ basis morphs into $\psi=C\psi^*=\tilde{\psi}$, with $C$ being given by the basis change matrix relating the $SO(3)$ representation's generator basis and the $T_3$-diagonal generator basis. There is a third set of generators of interest to us; the similarity transformation that rotated the $SO(3)$ basis to the cartesian one acted only on the matrix indices of the generators, not the triplet $a$ index in $T^a$. We can then use the rotation matrix in Eq.\ (\ref{eq:sphericalTransformation}) to rotate this remaining cartesian index to the spherical basis. It will transform in the same way the vector components did, as in Eq.\ (\ref{eq:sphericalComponentsTransformation}). We get
\begin{align}
    T_1&=-\frac{1}{\sqrt{2}}(T_x+iT_y)\;,\nonumber\\ 
    T_0&=T_3\;,\nonumber\\ 
    T_{-1}&=\frac{1}{\sqrt{2}}(T_x-iT_y)\;.   
\end{align}
This basis of generators is generally known as the spherical basis, same as the vector basis within the rotation matrix. For the doublet and triplet representations, the spherical generator bases are
\begin{align}\label{eq:generatorsSpherical}
    &\tau^1 = -\frac{1}{\sqrt{2}}\left(\begin{array}{cc}
        0 & 1 \\
        0 & 0
    \end{array}\right)\;,&&\tau^0 = \frac{1}{2}\left(\begin{array}{cc}
        1 & 0 \\
        0 & -1
    \end{array}\right)\;,&&\tau^{-1} = \frac{1}{\sqrt{2}}\left(\begin{array}{cc}
        0 & 0 \\
        1 & 0
    \end{array}\right)\;,\\[5mm]
    &t^1 = \left(\begin{array}{ccc}
        0 & -1 & 0 \\
        0 & 0 & -1 \\
        0 & 0 & 0
    \end{array}\right)\;,&& t^0 = \left(\begin{array}{ccc}
        1 & 0 & 0 \\
        0 & 0 & 0 \\
        0 & 0 & -1
    \end{array}\right)\;,&& t^{-1} = \left(\begin{array}{ccc}
        0 & 0 & 0 \\
        1 & 0 & 0 \\
        0 & 1 & 0
    \end{array}\right)\;.\nonumber
\end{align}
Having fully rotated all generator indices into the spherical basis allows us to build higher $SU(2)$ representations by contracting multiplets with the matrix and triplet generator indices in $(T^a)_{ij}$. As an example, it is well known that the $SU(2)$ product of two doublets decomposes into the triplet and singlet representations like $\mathbf{2}\times\mathbf{2}=\mathbf{3}\oplus\mathbf{1}$. When multiplying two doublets $\phi_i$ and $\eta_j$, we get the singlet representation by doing the full contraction $\phi_i\,\eta_i$, so that no indices remain. The triplet is built by contracting with the spherical generators' matrix indices $\phi_i\,(\tau^a)_{ij}\,\eta_j$, leaving the triplet index uncontracted so that the remaining representation is a triplet. 

\bibliographystyle{utphys}
\bibliography{mybib.bib}

\end{document}